\documentclass[twoside,twocolumn,9pt]{article}
\usepackage{extsizes}
\usepackage[super,sort&compress,comma,numbers]{natbib} 
\usepackage[usetitle]{rsc}
\usepackage[version=3]{mhchem}
\usepackage[left=1.5cm, right=1.5cm, top=1.785cm, bottom=2.0cm]{geometry}
\usepackage{balance}
\usepackage{times, mathptmx}
\usepackage{sectsty}
\usepackage{graphicx}
\usepackage{lastpage}
\usepackage[format=plain,justification=justified,singlelinecheck=false,font={stretch=1.125,small,sf},labelfont=bf,labelsep=space]{caption}
\usepackage{float}
\usepackage{fancyhdr}
\usepackage{fnpos}
\usepackage[english]{babel}
\addto{\captionsenglish}{%
  
}
\usepackage{array}
\usepackage{droidsans}
\usepackage{charter}
\usepackage[T1]{fontenc}
\usepackage[usenames,dvipsnames,table]{xcolor}
\usepackage{setspace}
\usepackage[compact]{titlesec}
\usepackage{hyperref}
\usepackage{tabularx}
\usepackage{dblfloatfix}

\usepackage{epstopdf}
\usepackage{blindtext}

\definecolor{cream}{RGB}{222,217,201}

\begin{document}

\pagestyle{fancy}
\thispagestyle{plain}
\fancypagestyle{plain}{
\renewcommand{\headrulewidth}{0pt}
}

\makeFNbottom
\makeatletter
\renewcommand\LARGE{\@setfontsize\LARGE{15pt}{17}}
\renewcommand\Large{\@setfontsize\Large{12pt}{14}}
\renewcommand\large{\@setfontsize\large{10pt}{12}}
\renewcommand\footnotesize{\@setfontsize\footnotesize{7pt}{10}}
\makeatother

\renewcommand{\thefootnote}{\fnsymbol{footnote}}
\renewcommand\footnoterule{\vspace*{1pt}%
\color{cream}\hrule width 3.5in height 0.4pt \color{black}\vspace*{5pt}} 
\setcounter{secnumdepth}{5}

\makeatletter 
\renewcommand\@biblabel[1]{#1}            
\renewcommand\@makefntext[1]%
{\noindent\makebox[0pt][r]{\@thefnmark\,}#1}
\makeatother 
\renewcommand{\figurename}{\small{Fig.}~}
\sectionfont{\sffamily\Large}
\subsectionfont{\normalsize}
\subsubsectionfont{\bf}
\setstretch{1.125} 
\setlength{\skip\footins}{0.8cm}
\setlength{\footnotesep}{0.25cm}
\setlength{\jot}{10pt}
\titlespacing*{\section}{0pt}{4pt}{4pt}
\titlespacing*{\subsection}{0pt}{15pt}{1pt}

\fancyfoot{}
\fancyfoot[LO,RE]{\vspace{-7.1pt}\includegraphics[height=9pt]{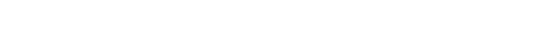}}
\fancyfoot[CO]{\vspace{-7.1pt}\hspace{13.2cm}\includegraphics{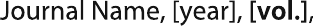}}
\fancyfoot[CE]{\vspace{-7.2pt}\hspace{-14.2cm}\includegraphics{head_foot/RF}}
\fancyfoot[RO]{\footnotesize{\sffamily{1--\pageref{LastPage} ~\textbar  \hspace{2pt}\thepage}}}
\fancyfoot[LE]{\footnotesize{\sffamily{\thepage~\textbar\hspace{3.45cm} 1--\pageref{LastPage}}}}
\fancyhead{}
\renewcommand{\headrulewidth}{0pt} 
\renewcommand{\footrulewidth}{0pt}
\setlength{\arrayrulewidth}{1pt}
\setlength{\columnsep}{6.5mm}
\setlength\bibsep{1pt}

\makeatletter 
\newlength{\figrulesep} 
\setlength{\figrulesep}{0.5\textfloatsep} 

\newcommand{\topfigrule}{\vspace*{-1pt}%
\noindent{\color{cream}\rule[-\figrulesep]{\columnwidth}{1.5pt}} }

\newcommand{\botfigrule}{\vspace*{-2pt}%
\noindent{\color{cream}\rule[\figrulesep]{\columnwidth}{1.5pt}} }

\newcommand{\dblfigrule}{\vspace*{-1pt}%
\noindent{\color{cream}\rule[-\figrulesep]{\textwidth}{1.5pt}} }

\makeatother

\twocolumn[
  \begin{@twocolumnfalse}
{\includegraphics[height=30pt]{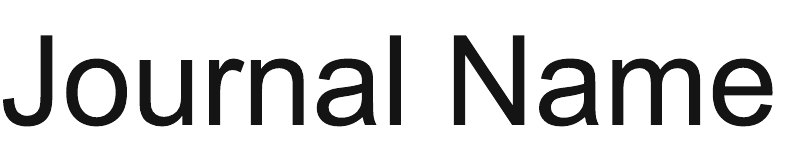}\hfill\raisebox{0pt}[0pt][0pt]{\includegraphics[height=55pt]{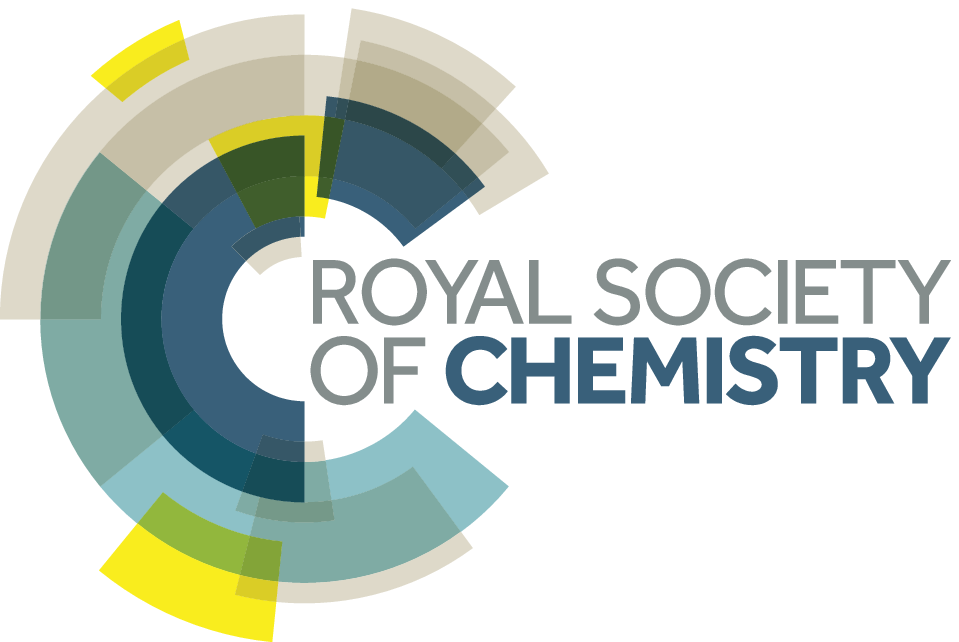}}\\[1ex]
\includegraphics[width=18.5cm]{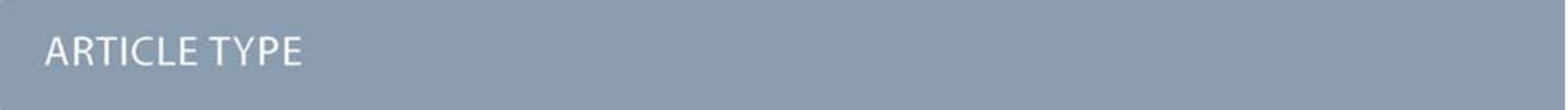}}\par
\vspace{1em}
\sffamily
\begin{tabular}{m{4.5cm} p{13.5cm} }

\includegraphics{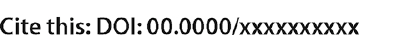} & \noindent\LARGE{\textbf{Laser-Annealing and Solid-Phase Epitaxy of Selenium Thin-Film Solar Cells}} \\
\vspace{0.3cm} & \vspace{0.3cm} \\

 & \noindent\large{Rasmus Nielsen,$^{\ast}$\textit{$^{a}$} Tobias H. Hemmingsen,\textit{$^{a}$} Tobias G. Bonczyk,\textit{$^{a}$} Ole Hansen,\textit{$^{b}$} Ib Chorkendorff,\textit{$^{a}$} and Peter C. K. Vesborg\textit{$^{a}$}} \\

\includegraphics{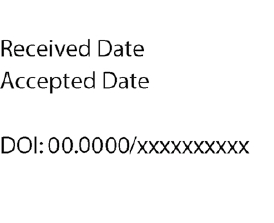} & \noindent\normalsize{Selenium has resurged as a promising photovoltaic material in solar cell research due to its wide direct bandgap of 1.95 eV, making it a suitable candidate for a top cell in tandem photovoltaic devices. However, the optoelectronic quality of selenium thin-films has been identified as a key bottleneck for realizing high-efficiency selenium solar cells. In this study, we present a novel approach for crystallizing selenium thin-films using laser-annealing as an alternative to the conventionally used thermal annealing strategy. By laser-annealing through a semitransparent substrate, a buried layer of high-quality selenium crystallites is formed and used as a growth template for solid-phase epitaxy. The resulting selenium thin-films feature larger and more preferentially oriented grains with a negligible surface roughness in comparison to thermally annealed selenium thin-films. We fabricate photovoltaic devices using this strategy, and demonstrate a record ideality factor of n=1.37, a record fill factor of FF=63.7\%, and a power conversion efficiency of PCE=5.0\%. The presented laser-annealing strategy is universally applicable and is a promising approach for crystallizing a wide range of photovoltaic materials where high temperatures are needed while maintaining a low substrate temperature.}




\end{tabular}

 \end{@twocolumnfalse} \vspace{0.6cm}

  ]

\renewcommand*\rmdefault{bch}\normalfont\upshape
\rmfamily
\section*{}
\vspace{-1cm}


\footnotetext{\textit{$^{a}$~SurfCat, DTU Physics, Technical University of Denmark, DK-2800 Kgs. Lyngby, Denmark.}}
\footnotetext{\textit{$^{b}$~DTU Nanolab, National Center for Nano Fabrication and Characterization, Technical University of Denmark, DK-2800, Kgs. Lyngby, Denmark.}}
\footnotetext{\textit{$^{\ast}$~ E-mail: raniel@dtu.dk}}

\footnotetext{\dag~Electronic Supplementary Information (ESI) available: Additional figures. See DOI: 00.0000/xxxxxxxxxx/}




\section*{INTRODUCTION}

Photovoltaics (PV) is considered an attractive renewable energy technology of key importance in the mitigation of climate change\cite{creutzig2017a}. While silicon solar cells have been dominating the global PV market for decades, a family of inorganic thin-film PV materials has emerged in recent years\cite{lee2015a, zakutayev2021a}. Thin-film PV offers a compelling pathway towards cost-competitive alternatives to silicon solar cells, as well as complementary materials for use in tandem devices\cite{mitzi2022a, liu2020a}. Among these emerging inorganic materials, selenium is a particularly promising candidate due to its single element composition, low melting point, and direct bandgap between 1.8 and 2.0 eV making it suitable for integration with lower bandgap photovoltaic devices\cite{lu2022a, zhu2019a, youngman2021a, nielsen2021a}. However, despite recent advances in the device architecture improving the power conversion efficiency (PCE) of selenium thin-film solar cells from 5.0\% to 6.5\%\cite{todorov2017a}, the PCE is still too low for tandem integration to be viable.

The key factors influencing the device performance of state-of-the-art selenium solar cells have been identified as the crystallinity and the morphology of the photoabsorber\cite{hadar2019b}. Typically, this layer is fabricated by evaporating an amorphous layer of selenium, followed by a crystallization process to transform the as-deposited film into the desired trigonal phase. The crystallization step is almost exclusively carried out by annealing the structure at temperatures of $\sim$200$^\circ$C on a pre-heated hotplate in air, but it is challenging to precisely control crystal growth using this strategy, and the optimal temperature is essentially a compromise between the degree of crystallinity and the surface morphology. Therefore, crystallizing the selenium thin-films using thermal annealing may be a bottleneck for realizing higher efficiency PV devices, and alternative approaches should be investigated.

\begin{figure*}[b!]
    \centering
    \vspace{0.25cm}
    \includegraphics[width=\textwidth,trim={0 0 0 0},clip]{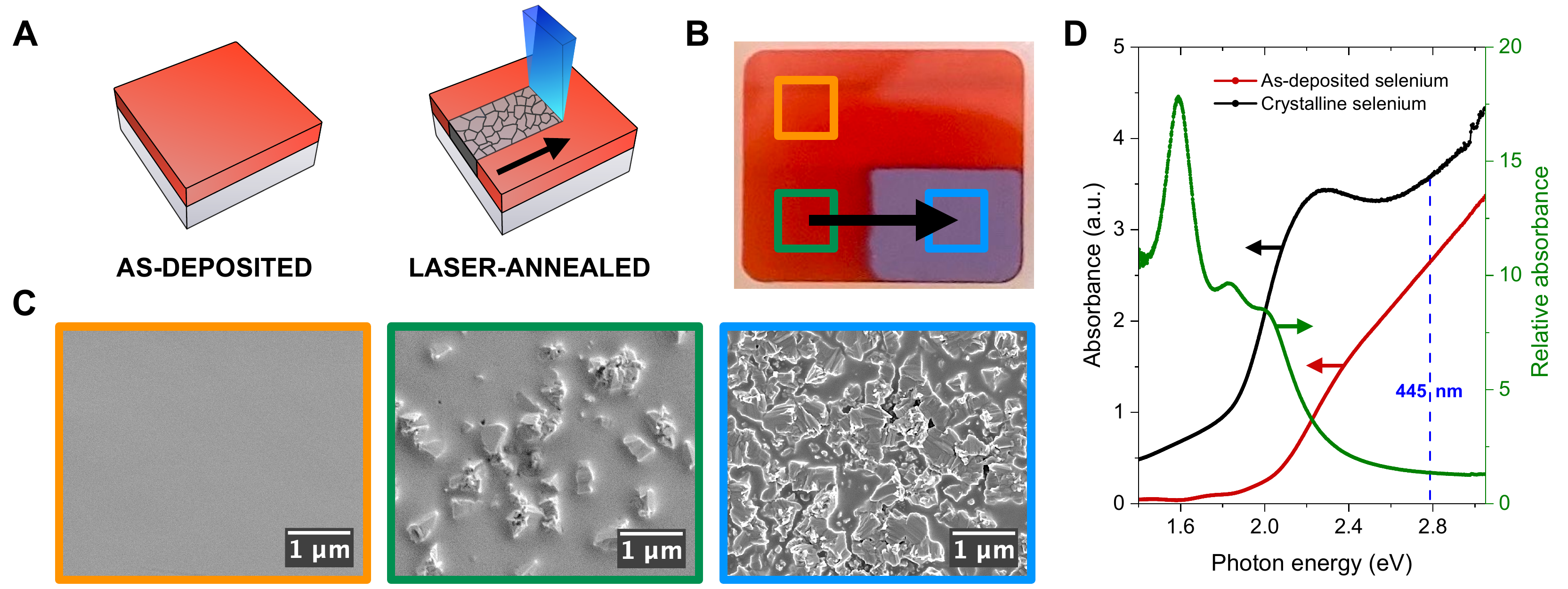}
    \caption{Comparison of the as-deposited and the laser-annealed surface. (\textbf{A}) Schematics illustrating the crystallization of the amorphous film induced by scanning the laser beam across the surface of the sample. (\textbf{B}) Optical and, (\textbf{C}) top-view SEM-images of 3 areas on the surface of the selenium thin-film. The area in the orange square has not been exposed to the laser, the area in the green square has been exposed during the first half of the scan, and the area in the blue square has been exposed during the second half of the scan. (\textbf{D}) The absorbance of a 300 nm selenium thin-film in the (as-deposited) amorphous and (thermally) crystalline phase.}
    \label{fig:Figure1}
\end{figure*}

Laser-annealing is a compelling alternative to thermal annealing, as the energy induced upon exposure may be accurately localized in time and space\cite{palneedi2018a}. This enables excellent control of the crystallization process as well as the crystal growth direction. The latter is particularly important in the case of trigonal selenium, as the optoelectronic material properties are anisotropic\cite{OriginOfPVLosses}. The crystallographic orientation of the grains in the selenium thin-film is therefore expected to affect the PV device performance. The crystallization of amorphous selenium under the influence of light was first studied half a century ago, where it was observed that illumination increases the growth rate of crystallites and modifies their external morphology\cite{dresner1968a,clement1974a}. The proposed mechanism suggests that the growth rate is determined by the flux of holes towards the grain boundary, and that the crystallites formed act as a sink for holes resulting from the topography of the energy bands. In a more recent study by Hadar \textit{et al.}\cite{hadar2019b}, the effect of light during the crystallization of selenium was investigated in the context of photovoltaic devices. By combining thermal annealing with strong white light illumination, they observed larger crystal grains and a reduction in the surface roughness. Furthermore, a significant improvement to the fill-factor (FF) was reported, indicating that the charge transport properties could be strongly affected by the non-equilibrium growth conditions induced by light.

In this work, we developed a novel laser-annealing strategy and report the first laser-annealed selenium thin-film solar cells. Unlike previous reports, our approach utilizes a monochromatic light source, which enhances the spatial confinement of the photo-generated carriers. Moreover, our laser-annealing strategy involves illuminating the photoabsorber through a semitransparent substrate, which, to the best of our knowledge, has never been reported before for any other material. As a result, a buried crystalline seed-layer is formed, serving as a growth template for solid-phase epitaxy. Compared to the thermally annealed control samples, the laser-annealed selenium thin-films exhibit negligible surface roughness and the presence of larger, more preferentially oriented grains. Our photovoltaic devices demonstrate a record fill factor FF=63.7\%, a record ideality factor n=1.37, and a power conversion efficiency of 5.0\%. The two-step crystallization process presented here is proposed as a universally applicable approach for achieving high-quality crystalline thin-films of any photovoltaic material synthesized on a substrate relevant for photovoltaic devices.

\section*{EXPERIMENTAL DETAILS}

FTO-coated glass substrates are ultrasonically cleaned in Milli-Q water, acetone and isopropanol, and dried using a nitrogen gun. The clean substrates are placed in a shadow mask and introduced into a commercial magnetron sputter chamber with a base pressure of $\textit{P}_{\text{base}}\sim10^{-7}$ mbar. Here, the samples are sputter cleaned in 20 mTorr of Ar for 60 seconds, followed by the deposition of a $\sim45$ nm layer of TiO$_\text{2}$. The TiO$_\text{2}$ thin-film is formed by sputtering metallic Ti in a reactive atmosphere of Ar/O$_\text{2}$ (60/6 sccm) at 5 mTorr with an elevated substrate temperature of 400$^\circ$C. To further improve the crystallinity and increase the density of oxygen vacancies, the samples are subsequently annealed at 500$^\circ$C for 1 hour under high vacuum. Oxygen vacancies $V_\text{O}^{+2}$ are known to act as native n-type dopants in TiO$_\text{2}$, thus improving the transport of electrons through the selective contact.

\begin{figure*}[b!]
    \centering
    \vspace{0.4cm}
    \includegraphics[width=\textwidth,trim={0 0 0 0},clip]{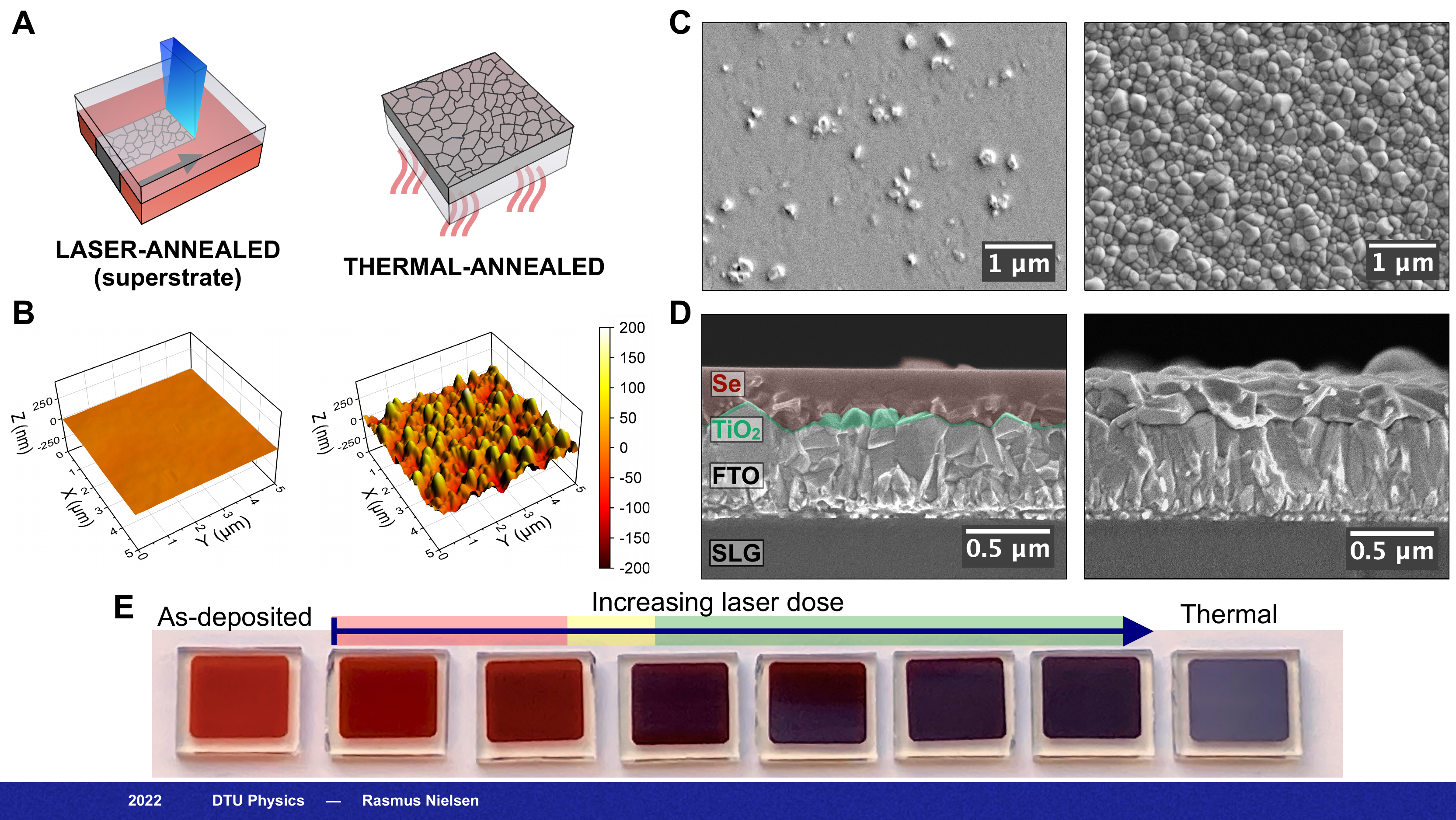}
    \vspace{0.2cm}
    \caption{Comparing the superstrate laser-annealing strategy to the standard thermal annealing strategy. (\textbf{A}) Schematics illustrating how energy is induced into the selenium thin-film by either lasing or heating through the substrate. (\textbf{B}) AFM-images and, (\textbf{C}) top-view SEM-images of the selenium thin-film surfaces after the two respective processes, complemented by (\textbf{D}) cross-sectional SEM-images. At this processing step the device structure comprise SLG/FTO/TiO$_\text{2}$/Te/poly-Se. (\textbf{E}) Optical image of the selenium thin-film surface as a function of increasing laser dose.}
    \label{fig:Figure2}
\end{figure*}

The samples are left to cool down to room temperature under high vacuum, and then briefly exposed to air in the transfer to a custom-built thermal evaporator with a base pressure of $\textit{P}_{\text{base}}\sim10^{-8}$ mbar. Here, a thin layer of $\approx$1 nm of Te deposited, followed by the deposition of $\approx$300 nm of Se. The purpose of the ultra-thin Te layer is to improve adhesion and uniformity of the selenium thin-film during the crystallization process\cite{nakada1985a}. The batch of reference samples are crystallized by thermal annealing in air at 190$^\circ$C for 4 minutes in a pre-heated aluminium oven. The other samples are placed in a 3D-printed sample holder facing a 445 nm continuous-wave diode laser. The laser beam is focused to a diameter of $\textit{d} \approx$1.75 mm, and the optical power density is adjusted to $\textit{P}\approx$100 W/cm$^2$. The laser beam is scanned in a serpent-like raster pattern across the surface of the samples using a set of stepper motors (step size and line separation of 40 $\mu$m), and the total laser dose is varied by adjusting the raster speed. After the crystallization step, $\sim$15 nm of MoO$_\text{x}$ is sputter deposited from a ceramic MoO$_\text{3}$ target in a reactive atmosphere of Ar/O$_\text{2}$ (60/0.6 sccm) at 5 mTorr. Finally, Au-contacts are sputter deposited on the samples to complete a device structure comprising glass/FTO/TiO$_\text{2}$/Te/Se/MoO$_\text{x}$/Au. The experimental details have been elaborated in the Supplementary Information, along with descriptions of the various characterization methods.\\

\section*{RESULTS AND DISCUSSION}

The initial laser-annealing experiments were inspired by the novel approach developed by You et al. for crystallizing perovskite thin-film photoabsorbers\cite{you2020a}, and have been schematically illustrated in Figure \ref{fig:Figure1}A. A laser spot is scanned across the surface of the as-deposited film, transforming the layer from an amorphous into the desired crystalline phase in the wake of the beam. Figure \ref{fig:Figure1}B shows an optical image of a 12 mm x 10 mm x 300 nm selenium thin-film, where the top half of the sample has been left unexposed as a reference of the as-deposited film. The bottom half of the sample has been laser-annealed by scanning the laser beam in a raster pattern initiated on the left-hand side. From visual inspection, the surface of the sample is divided into three distinctive areas; the as-deposited film (orange), the film exposed during the first half of the laser scan (green), and the film exposed during the second half of the laser scan (blue). The three regions have been examined using top-view SEM as shown in Figure \ref{fig:Figure1}C. In the first half of the scan, large crystallites are observed in the surface of the film, and the coexistence of the amorphous and crystalline (trigonal) phases is confirmed using GIXRD (ESI, Fig S.6). However, as the scan progress, the morphology changes and the integrity of the surface is compromised by pinholes. As the optical power density (100 W/cm$^\text{2}$) and the raster speed (12.5 mm/s) remain constant during the scan, the observed morphological changes cannot be attributed to a variation in the laser dose. Another possible explanation is a thermal runaway, resulting from a gradual rise in the temperature of the sample surface. Considering the influence of temperature and crystallinity on absorptivity, it is conceivable that the optical energy is absorbed within a progressively thinner surface layer as the scan proceeds. Consequently, the surface temperature may rise to a point where selenium starts to re-evaporate, leading to the formation of voids and pinholes.

\begin{figure*}[ht!]
    \centering
    \includegraphics[width=\textwidth]{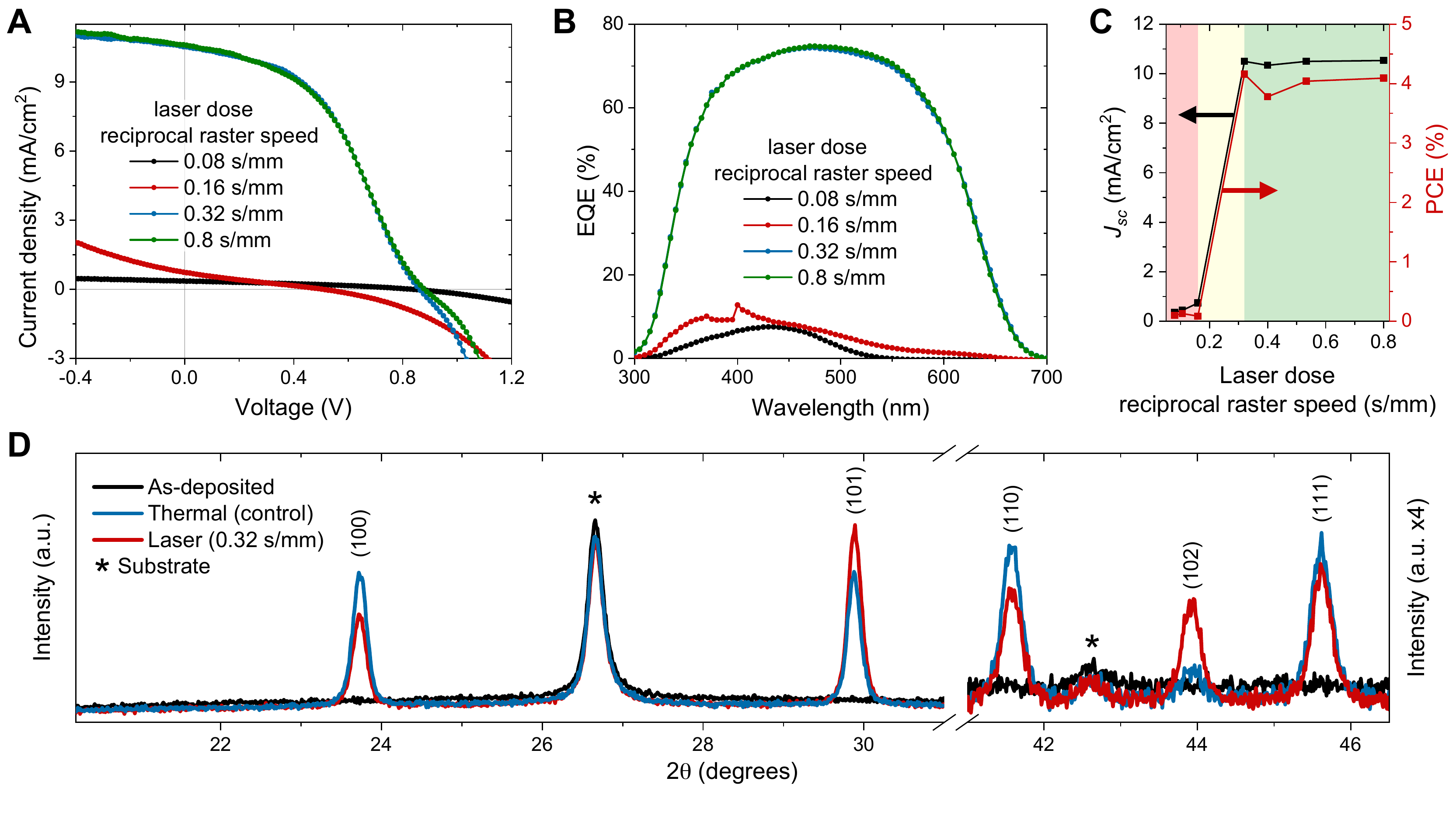}
    \caption{Photovoltaic device performance as a function of increasing laser dose. (\textbf{A}) Current-voltage (\textit{J-V}) curves measured under 1 Sun conditions. (\textbf{B}) External quantum efficiency (EQE) spectra measured under short-circuit conditions. (\textbf{C}) Performance metrics as a function of increasing laser dose. (\textbf{D}) XRD patterns (Cu K$\alpha$) of a selenium thin-film crystallized using superstrate laser-annealing (0.32 s/mm) compared to a thermally annealed control sample. Peaks marked with an asterisk are attributed to the FTO-coated glass substrate.}
    \label{fig:Figure3}
\end{figure*}

To gain a deeper understanding of the optical dispersion of the laser within the film, we measured UV-visible absorbance spectra of 300 nm amorphous and crystalline selenium thin-films synthesized on quartz substrates, as shown in Figure \ref{fig:Figure1}D. The relative absorbance between the two phases of the 445 nm laser is 1.35, and the penetration depth $\delta =1/\alpha\left(\lambda\right)$ is 38 nm and 49 nm in the crystalline and amorphous phase, respectively. These findings suggest that the optical energy of the laser is primarily absorbed in close proximity to the surface of the film, and becomes even more localized once large crystallites have been formed. Due to the higher absorbance of the crystalline phase, the temperature of the crystallites is expected to surpass that of the surrounding amorphous material. If thermal gradients across the grain boundaries promote crystal growth, opting for a longer wavelength laser to further increase the relative absorbance could potentially lead to the formation of even larger grains. Nevertheless, due to the high vapor pressure of selenium, the parameter space for achieving a void- and pinhole-free crystalline thin-film through laser annealing of the surface is exceedingly narrow and not necessarily consistent throughout the scan.

Seeing that it was not possible to fabricate functional selenium solar cells using the aforementioned laser-annealing strategy, an alternative approach was developed, as schematically illustrated in Figure \ref{fig:Figure2}A. By taking advantage of the transparency of the SLG/FTO/TiO$_\text{2}$ substrate to the 445 nm laser, laser-annealing of the amorphous selenium thin-film was performed through the substrate. Using this "\textit{superstrate}" laser-annealing strategy, the optical energy is predominantly absorbed in the vicinity of the carrier separating junction instead of the open surface from where selenium can re-evaporate away.

Figure \ref{fig:Figure2}E displays an optical image of the selenium thin-films after laser-annealing through the substrate, in comparison to the as-deposited and thermally annealed films. The left-hand sample represents the first superstrate laser-annealed sample, receiving the same laser dose as the sample in Figure \ref{fig:Figure1}B. However, as only a slight darkening of the film was observed, the raster speed was successively halved for each subsequent sample until the dose had increased by a factor of 10. The visual appearance of the superstrate laser-annealed samples significantly differs from that of both the substrate laser-annealed and the thermally annealed samples. Instead of a grey appearance, the color is a dark purple/black shade, which may be attributed to distinct surface morphology differences observed through AFM and top-view SEM in Figure \ref{fig:Figure2}B and C, respectively. The root-mean-square roughness was quantified over a  $\mu$m$^\text{2}$ area, measuring 4.2 nm and 44.4 nm on the laser-annealed and thermally annealed samples, respectively. Top-view SEM-images revealed only slight penetration of the crystallites into the surface of the laser-annealed sample, in contrast to the thermally annealed sample, where large, bulky grains and their boundaries are clearly visible. Moreover, significant differences in the external morphology of the grains in the laser-annealed and thermally annealed samples are observed, consistent with previous findings reported by Dresner and Stringfellow\cite{dresner1968a}. To gain further insights, cross-sectional SEM-images are acquired as shown in Figure \ref{fig:Figure2}D, revealing the formation of selenium crystallites buried under a still amorphous layer of selenium. Additional top-view and cross-sectional SEM-images can be found in the Supplementary Information.

\begin{figure*}[ht!]
    \centering
    \includegraphics[width=\textwidth]{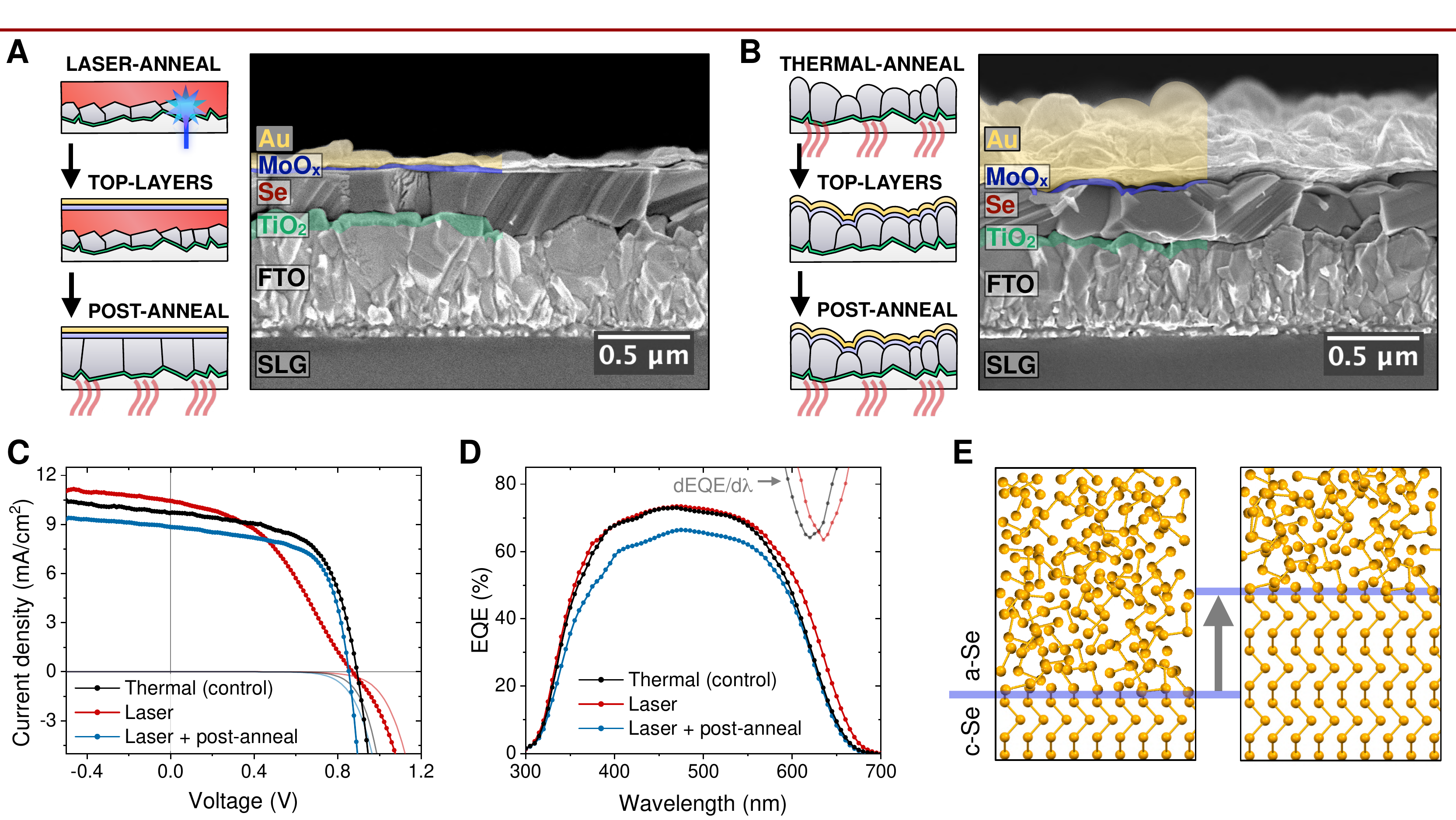}
    \caption{Investigating the effect of a final post-annealing processing step. (\textbf{A}) Process flow and a cross-sectional SEM-image of a laser-annealed device. (\textbf{B}) Same as (A) for a thermally annealed control device. (\textbf{C}) Current-voltage (\textit{J-V}) curves measured under 1 Sun conditions. (\textbf{D}) External quantum efficiency (EQE) spectra measured under short-circuit conditions. The first derivatives of the onsets have been included to visualize the shift of the inflection point. (\textbf{E}) Schematic illustration of solid-phase epitaxial (SPE) growth of crystalline selenium thin-films.}
    \label{fig:Figure4}
\end{figure*}

Following the superstrate laser-annealing step, MoO$_\text{x}$ and Au are sputter deposited to finalize a device structure comprising SLG/FTO/TiO$_\text{2}$/Te/poly-Se/a-Se/MoO$_\text{x}$/Au. We examine the device performance with varying laser doses, quantifying the dose using the reciprocal raster speed while maintaining a fixed optical power density of 100 W/cm$^2$. Figure \ref{fig:Figure3}A and B shows current-voltage (\textit{J-V}) and external quantum efficiency (EQE) measurements. We observe a threshold for the laser dose required for the device to function, with no evident effect on PV performance upon exceeding this threshold. This observation has been illustrated in Figure \ref{fig:Figure3}C. It is noteworthy that the selenium thin-film undergoes a visual transformation exactly at this threshold, as seen in Figure \ref{fig:Figure2}E. This suggests a significant change in the degree of crystallinity. However, despite achieving functioning devices with power conversion efficiencies of PCE=4\%, \textit{"S-kinks"} are observed in the \textit{J-V}--curves near the open-circuit voltage. S-kinks are often reported in the context of emerging photovoltaic devices, arising from bias-dependent charge transport barriers at one of the carrier-selective contacts\cite{saive2019a}. As previously observed from the cross-sectional SEM-images in Figure \ref{fig:Figure2}D, the large selenium crystallites are buried under a layer of still amorphous selenium. As amorphous selenium is known to be highly resistive, this could pose a critical challenge for efficient charge carrier extraction.
 
Charge transport in selenium thin-films is rather complex due to the trigonal symmetry of the crystalline phase, which leads to highly anisotropic carrier mobilities\cite{nielsen2021a}. Aligning the (001) plane parallel to the substrate, which corresponds to an arrangement of the helical chains along the charge transport direction, should improve charge transport and ultimately enhance device performance. To investigate the crystallographic orientation of the grains in the laser-annealed selenium thin-films, the XRD patterns shown in Figure \ref{fig:Figure3}D are recorded using Bragg-Brentano geometry. When comparing the laser-annealed sample to the thermally annealed control sample, a pronounced difference in the relative intensities of the (100) and (101) reflections is observed. This result indicates that a larger fraction of the grains in the laser-annealed sample exhibit a preferential orientation compared to those in the thermally annealed sample. As the optical energy of the laser is absorbed in the amorphous selenium thin-film, a thermal gradient is formed in the superstrate configuration, extending from the TiO$_\text{2}$/Se-interface to the open surface. This thermal gradient may potentially facilitate the preferential growth of the helical chains along its direction.

\begin{figure*}[b!]
    \centering
    \vspace{0.25cm}
    \includegraphics[width=\textwidth]{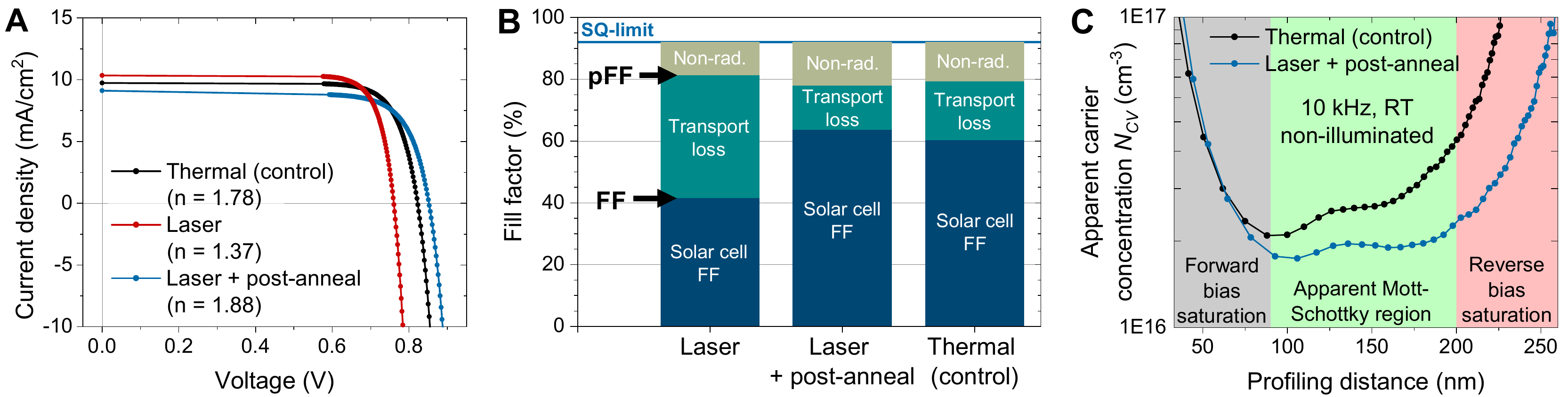}
    \caption{Analysis of carrier transport losses. (\textbf{A}) Current-voltage (\textit{J-V}) curves reconstructed from SunsVoc measurements. (\textbf{B}) Chart illustrating the loss mechanisms lowering the device FF below the SQ-limit. (\textbf{C}) Apparent carrier concentration as a function of profiling depth from C-V measurements.}
    \label{fig:Figure5}
\end{figure*}

In continuation of the promising results obtained through superstrate laser-annealing, an additional processing step is introduced. After the deposition of the top layers, the entire stack is thermally annealed at 190$^\circ$C for 4 min in air to facilitate the crystallization of the remaining amorphous layer of selenium, while maintaining the smooth morphology. This transformation is evident from the well-defined crystal facets observed in the cross-sectional SEM-image in Figure \ref{fig:Figure4}A. The resulting device stands in contrast to the control device shown in Figure \ref{fig:Figure4}B. Schematic diagrams illustrating the final processing steps have been included alongside the SEM-images to emphasize the crystallization step as the only difference between the two devices. To avoid confusion with the thermal annealing step used for our control devices, the final processing step is referred to as \textit{"post-annealing"}.

Figure \ref{fig:Figure4}C and D shows IV and EQE measurements of a superstrate laser-annealed device before and after post-annealing, in comparison to a thermally annealed control device. The apparent S-kink is eliminated upon post-crystallization of the amorphous layer of selenium; however, this improvement comes at the expense of a reduced short-circuit current density. This reduction is also observed in the EQE, showing a decrease in carrier collection across all wavelengths. These findings suggest that the amorphous selenium layer passivates the interface with the hole-selective MoO$_\text{x}$ contact, thereby reducing the interfacial recombination velocity. In contrast, the control device features no layer of amorphous selenium, and the EQE and short-circuit current density are similar to those of the laser-annealed sample before post-annealing. This result may be attributed to the formation of a passivating oxide layer when the control device is thermally annealed in air. Alternatively, as the transformation of amorphous to trigonal selenium involves a 10\% volume reduction\cite{grosse1978a}, the post-annealing process may impose strain or stress on the already deposited top layers in the case of the laser-annealed device. 

Another notable difference resulting from the post-annealing of the laser-annealed sample is a shift in the photovoltaic bandgap. The photovoltaic bandgap is determined from the inflection point of the EQE in Figure \ref{fig:Figure4}D. It has been predicted that strain has a pronounced influence on the bandgap of selenium\cite{singh2020a}. However, the fact that the inflection points of the control device and the laser-annealed device after post-annealing are identical suggests that other factors may be responsible for the observed shift. Another hypothesis pertains to the alloying of selenium and tellurium, which is known to decrease the bandgap of selenium\cite{hadar2019a}. In our devices, tellurium is present as a thin adhesion layer of approximately 1 nm. As a result of the more localized heating during laser-annealing, the bulk inter-diffusion of tellurium is likely confined to a smaller volume of the selenium thin-film compared to the thermally annealed control samples. Consequently, the local concentration of tellurium is anticipated to be higher in the selenium crystallites before the post-annealing process. During the post-annealing process, tellurium is expected to diffuse throughout the width of the photoabsorbing layer, leading to a more diluted local concentration and ultimately an effective increase in the bandgap. The presence of tellurium at the surface and throughout the photoabsorbing layer is confirmed using XPS depth profiling (Figure S.3).

The post-annealing process is depicted in Figure \ref{fig:Figure4}E, showing a crystalline seed layer in intimate contact with an amorphous layer of selenium. During the post-annealing, the c-Se/a-Se interface moves upwards in a process known as solid-phase epitaxy (SPE). Typically, SPE involves depositing an amorphous layer onto of a crystalline substrate, followed by a heating step to crystallize the deposited film using the substrate as a growth-template. However, our two-step crystallization process differs in that the seed layer is formed within the as-deposited film, making it applicable to other material systems synthesized on a substrate transparent to the laser wavelength. SPE has been extensively studied for the growth of c-Si and c-Ge, with reported growth temperatures as low as 500$^\circ$C and 350$^\circ$C for Si and Ge, respectively\cite{claverie2010a, nishinaga2014a}. These temperatures are significantly lower than those typically used for growing crystalline silicon and germanium. This implies that the optimal crystal growth temperature of selenium could be reduced by growing the crystalline thin-film from a buried seed layer. Such an approach would effectively mitigate the morphological challenges encountered in selenium thin-films crystallized at temperatures exceeding 180$^\circ$C\cite{hadar2019b}.

The post-annealing treatment effectively eliminated the charge transport limitations, leading to an impressive fill factor of FF=63.7\% and a power conversion efficiency of PCE=5.0\%. It is worth noting that this FF value is the highest reported among selenium solar cells, and that the FFs of the laser-annealed samples reproducibly outperform those of the thermally annealed control samples, regardless of the range of laser doses investigated (Figure S.9 and S.10). The improved FF could be attributed to the more optimally aligned crystal grains and the flat surface morphology. Previous studies have shown that selenium solar cells are highly sensitive to absorber thickness, with an optimal value of approximately 300 nm\cite{youngman2021a}. The flat morphology of the laser-annealed sample eliminates potential shunting pathways, such as pinholes, and reduces the local variations in the absorber thickness, leading to more spatially homogeneous devices.

Despite achieving a record fill factor with great reproducibility regardless of laser dose, the obtained value still falls significantly below the SQ-limit for a 1.95 eV bandgap material, which is FF$_\text{SQ}$=92.1\%. To investigate the origin of this significant fill factor deficit, we conducted SunsVoc measurements to reconstruct IV curves with no carrier transport losses, shown in Figure \ref{fig:Figure5}A. Remarkably, the device ideality factor for the laser-annealed sample was found to be n=1.37, which is significantly better than any other previously reported value in literature. However, the ideality factor worsened after the post-annealing, supporting the hypothesis that the amorphous selenium layer passivates the poly-Se/MoO$_\text{x}$-interface. A loss chart has been constructed in Figure \ref{fig:Figure5}B by combining the pseudo fill factors (pFF) obtained from SunsVoc measurements, where the effects of series resistance has been eliminated, with the actual device fill factors. This chart illustrates the residual losses that are not accounted for by series resistance, resulting from non-radiative recombination occurring at interfaces and within the bulk of the photoabsorber. Comparing the laser-annealed sample to the thermally annealed control sample, we observed lower non-radiative recombination losses prior to the post-annealing, with carrier transport losses being dominant. After the post-annealing process, the laser-annealed sample demonstrated a higher fill factor compared to the control sample. However, upon conducting a carrier transport loss analysis, it was determined that the improvement in fill factor was not attributed to lower non-radiative recombination losses. Instead, it is primarily due to a reduction in transport losses. As the highly resistive layer of amorphous selenium has been fully crystallized after this processing step, the reduced transport losses could be the result of an improved photoabsorber conductivity. Finally, capacitance-voltage measurements have been conducted to examine the apparent carrier concentration of both devices. Despite having a lower apparent carrier concentration, the laser-annealed device demonstrate fewer transport losses, which supports the hypothesis that the crystal grains making up the photoabsorbing layer are in fact more ideally aligned and better electrically contacted. 

An important distinction between laser-annealing and thermal annealing lies in the thermal temperature of the charge carriers within the material during the growth process. When photo-generated charge carriers split the Fermi-level into two quasi-Fermi levels, the electrons and holes are no longer in thermal equilibrium with their surroundings. This non-equilibrium state of the charge carriers has implications for the formation energies, and hence concentrations, of both extrinsic and intrinsic defects in the crystal\cite{kim2020a}. Therefore, laser-annealing has the potential to serve as a defect-engineering tool for non-equilibrium growth of high-quality crystalline thin-films, levering the chemical potential of charge carriers without introducing extrinsic dopants.\\

\section*{CONCLUSIONS}

In summary, we have introduced a novel laser-annealing strategy and successfully fabricated the first laser-annealed selenium thin-film solar cells. Our two-step crystallization approach involves laser-annealing through a device-relevant semitransparent substrate, resulting in the formation of buried selenium crystallites, serving as a growth template for solid-phase epitaxy. In comparison to thermally annealed control devices, our laser-annealed selenium thin-films exhibit notable improvements, including reduced surface roughness and the presence of larger, more preferentially oriented grains. The resulting photovoltaic devices demonstrate promising performance metrics, with a record fill factor of FF=63.7\%, a record ideality factor of n=1.37, and a power conversion efficiency of PCE=5.0\%.

This work underscores the potential of laser-annealing as a viable alternative to thermal annealing for enhancing the performance of selenium thin-film solar cells. Furthermore, the two-step crystallization process presented here showcases its universal applicability for achieving high-quality crystalline thin-films across various photovoltaic materials on substrates relevant to photovoltaic devices. With its inherent advantages of being fast, scalable, and energy-efficient, laser-annealing may pave the way for the development of more efficient and cost-effective thin-film solar cells based not only on selenium but also on other inorganic materials.\\

\section*{Conflicts of interest}
There are no conflicts to declare.\\

\section*{Acknowledgements}
This work was supported by the Independent Research Fund Denmark (DFF) grant 0217-00333B.\\



\balance


\bibliography{references} 
\bibliographystyle{rsc} 

\end{document}


\maketitle
\vspace{-1.5cm}
\section*{Experimental details}

\paragraph*{Materials.~~} Amorphous selenium (99.999+\%, metals basis) and tellurium (99.9999\%, metals basis) shots were purchased from Alfa Aesar. FTO-coated glass substrates ($\sim7\,\Omega/\text{sq}$) were purchased from Sigma-Aldrich. Ti (99.995\%), MoO$_\text{3}$ (99.9\%), and Au (99.99\%) sputtering targets were purchased from AJA International.

\paragraph*{Laser-annealing setup.~~} The laser-annealing setup is a repurposed laser engraver from NEYE Tools (KZ3000). The laser is a 445 nm continuous-wave diode laser, and the XY-movement is implemented using a set of stepper motors with worm drives. To enable microstepping, a set of DRV8825 drivers are integrated in the electronics control circuit and run using an Arduino Uno. To properly secure the samples and enable superstrate laser-annealing, a sample holder with a hollow center is 3D-printed, and the laser is finally focused using a focusing lens a distance $\sim$5 cm above the sample surface.

\paragraph*{Device characterization.~~} Current-voltage (\textit{J-V}) measurements of photovoltaic devices are measured using a Keithley 2561A source meter with 4-terminal sensing under 1 sun illumination (Newport 94082A solar simulator, class ABA, equipped with a 1600 W Xe arc lamp and appropriate AM1.5G filters). The light intensity is calibrated in the plane of the device under test using a reference solar cell from Orion. As no mask aperture is used during the acquisition, the active area is determined by calculating the AM1.5G equivalent current density using the integral of the external quantum efficiency (EQE) spectrum of the device under test. EQE spectra are measured using the QEXL from PV Measurements calibrating using a silicon reference photodiode. Suns-Voc measurements are acquired using the WCT-120 from Sinton Instruments.

\paragraph*{Other characterizations.~~} Scanning electron microscopy (SEM) images are acquired using a Supra 40 VP SEM from Zeiss. UV-vis absorbance spectra are measured at room temperature using a UV-2600 spectrophotometer from Shimadzu. XRD patterns are measured in a Panalaytical Empyrean XRD using a parallel plate collimator at an angle of 0.5$^\circ$ for doing GIXRD, whereas BraggBrentano measurements use a programmable divergence (anti-scatter) slit. In both configurations the source was a Cu LFF HR gun operated at 45 kV and 40 mA, with K$\alpha$1 = 1.540598 Å. Photoemission spectroscopy measurements are carried out in the commercial Thermofisher Scientific Nexsa XPS system. AFM imaging was performed in air using a FlexAFM (Nanosurf,CH) operated in tapping mode. A gold-coated probe (Tap150GD-G, BudgetSensors, BG) of 150 kHz and a spring constant of 5 N/m were used. All images (256 samples/line x 256 lines) were completed at room temperature. AFM derived raw data was leveled and the root-mean-square (RMS) roughness obtained with the scanning probe microscopy data analysis software Gwyddion. Dust or other airborne contamination on the surface of the freshly synthesized samples were removed by inert, pure compressed gas (RS Pro Air duster, RS Components, UK) prior AFM imaging.

\begin{figure*}[b!]
    \centering
    \includegraphics[width=0.4004\textwidth,trim={70 60 475 140},clip]{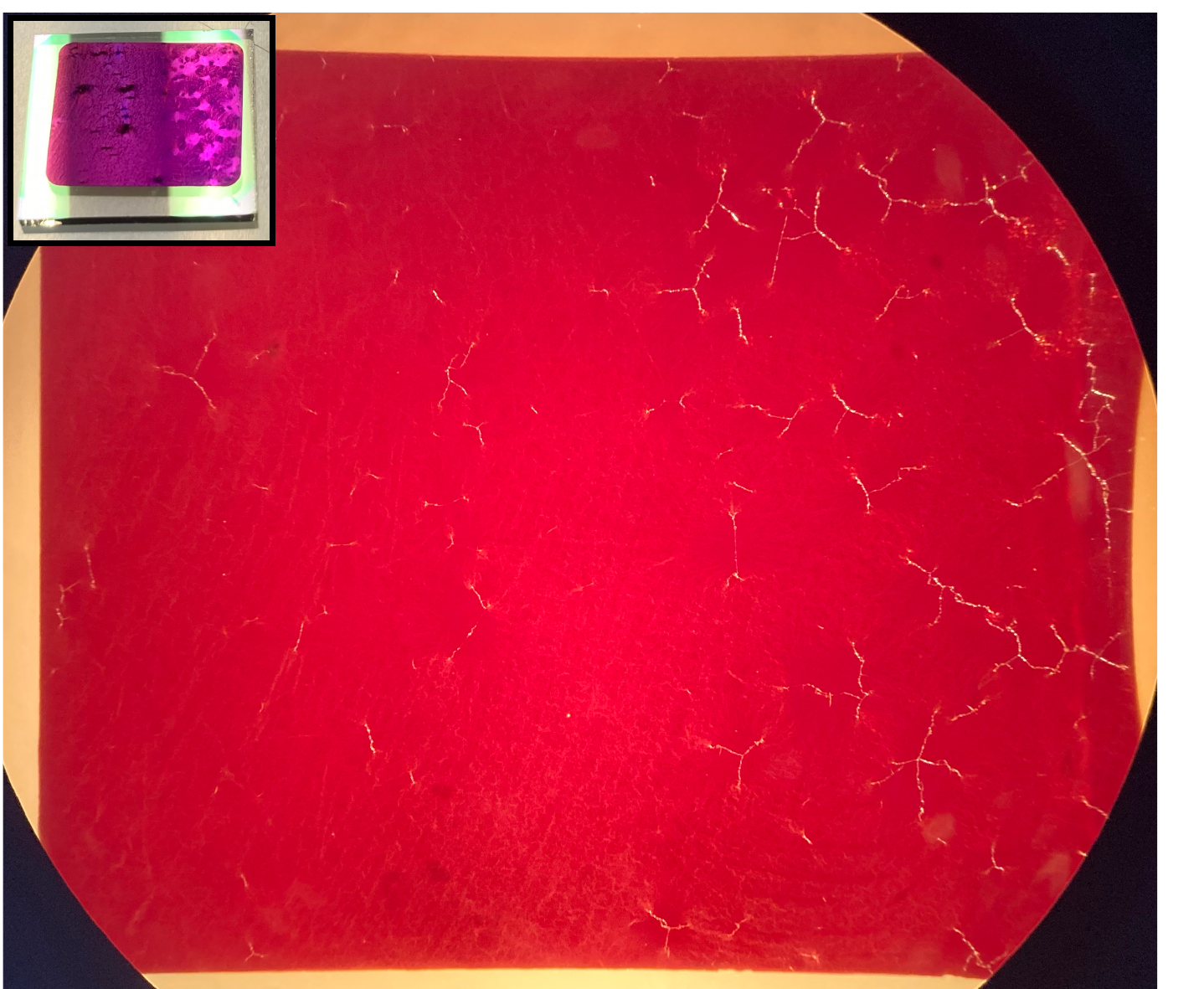}
    \quad
    \includegraphics[width=0.495\textwidth,trim={0 0 0 0},clip]{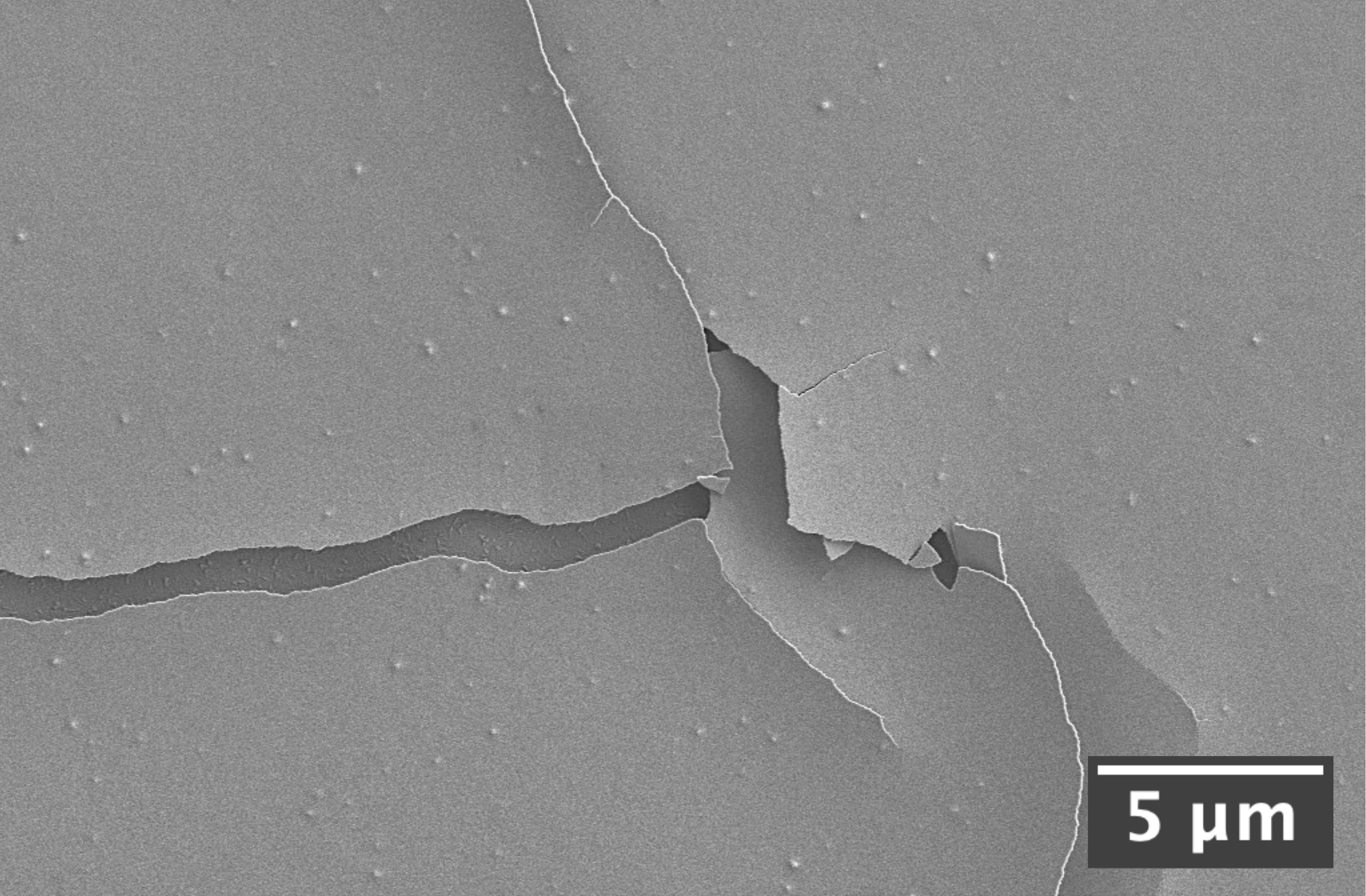}
    \caption{Attempting to thermally crystallize selenium thin-films after depositing MoO$_\text{x}$ on the as-deposited film. \textbf{Left:} Optical microscopy image of a 12 x 10 mm selenium thin-film and an inset of an optical photography of the sample showing the formation of large cracks. \textbf{Right:} Top-view SEM image of a crack.}
    \label{fig:ESI1}
\end{figure*}

\begin{figure*}[ht!]
    \centering
    \includegraphics[width=\textwidth,trim={50 0 50 0},clip]{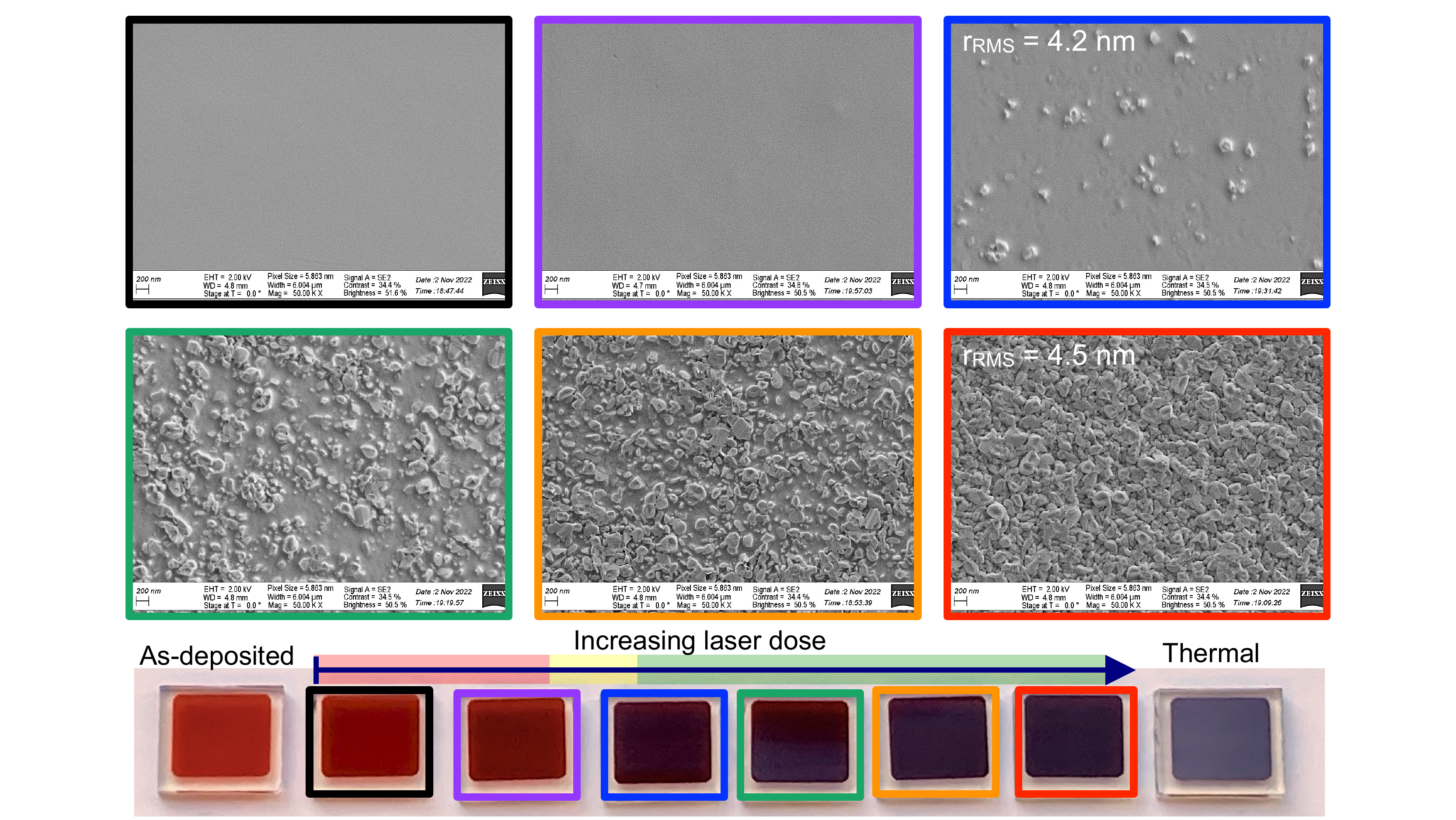}
    \caption{Top-view SEM images of selenium thin-films as a function of superstrate laser annealing dose. The root-mean-square roughness acquired from AFM-measurements are $\sim$5 nm even on samples receiving the highest laser doses.}
    \label{fig:ESI7}
\end{figure*}

\begin{figure*}[ht!]
    \centering
    \includegraphics[width=0.5\textwidth,trim={0 0 0 0},clip]{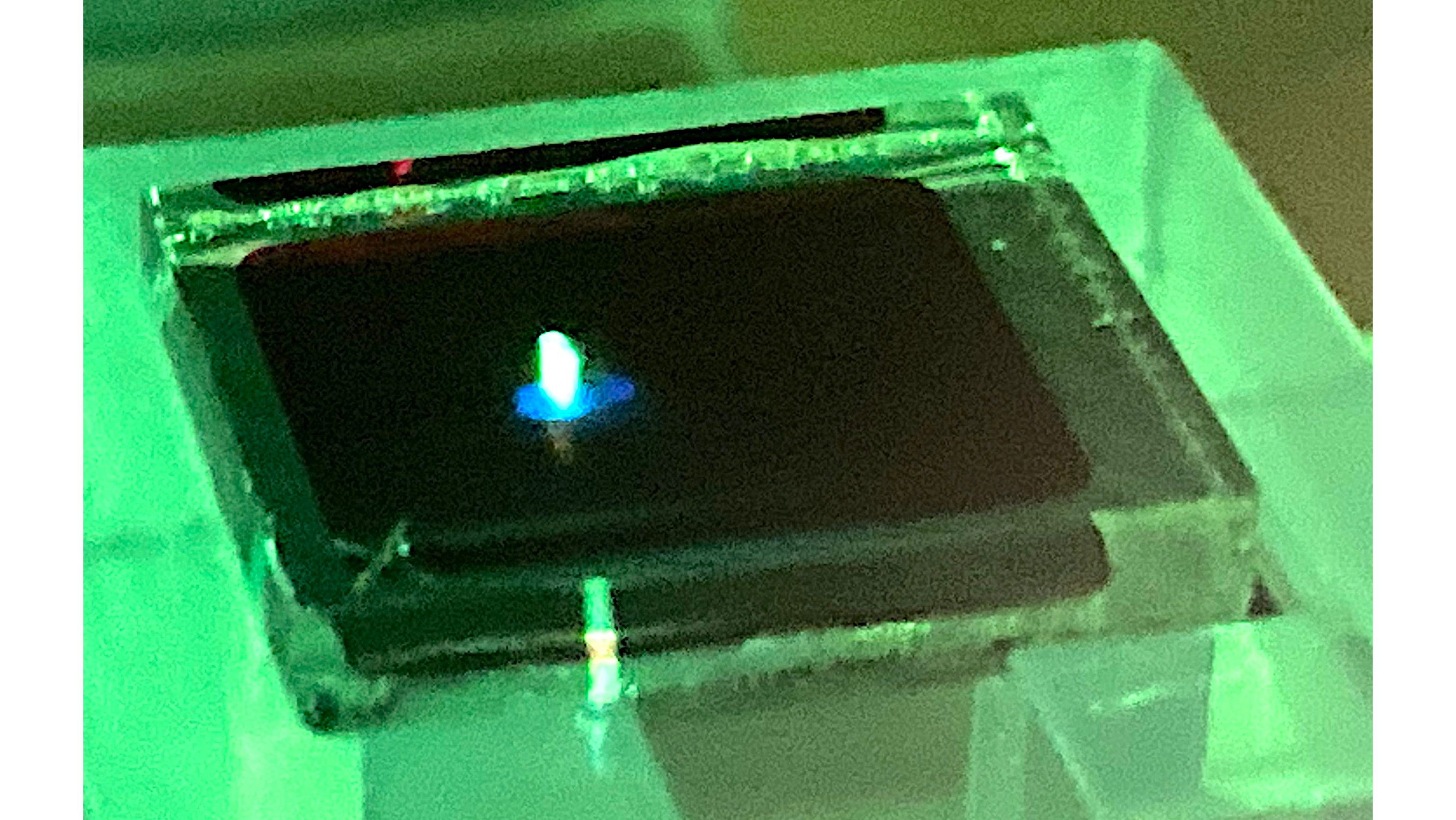}
    \caption{Optical image of superstrate laser-annealing, where the laser beam is observed to transmit through the semitransparent substrate and be absorbed in the amorphous selenium thin-film .}
    \label{fig:ESI10}
\end{figure*}

\begin{figure*}[ht!]
    \centering
    \includegraphics[width=\textwidth,trim={0 50 0 50},clip]{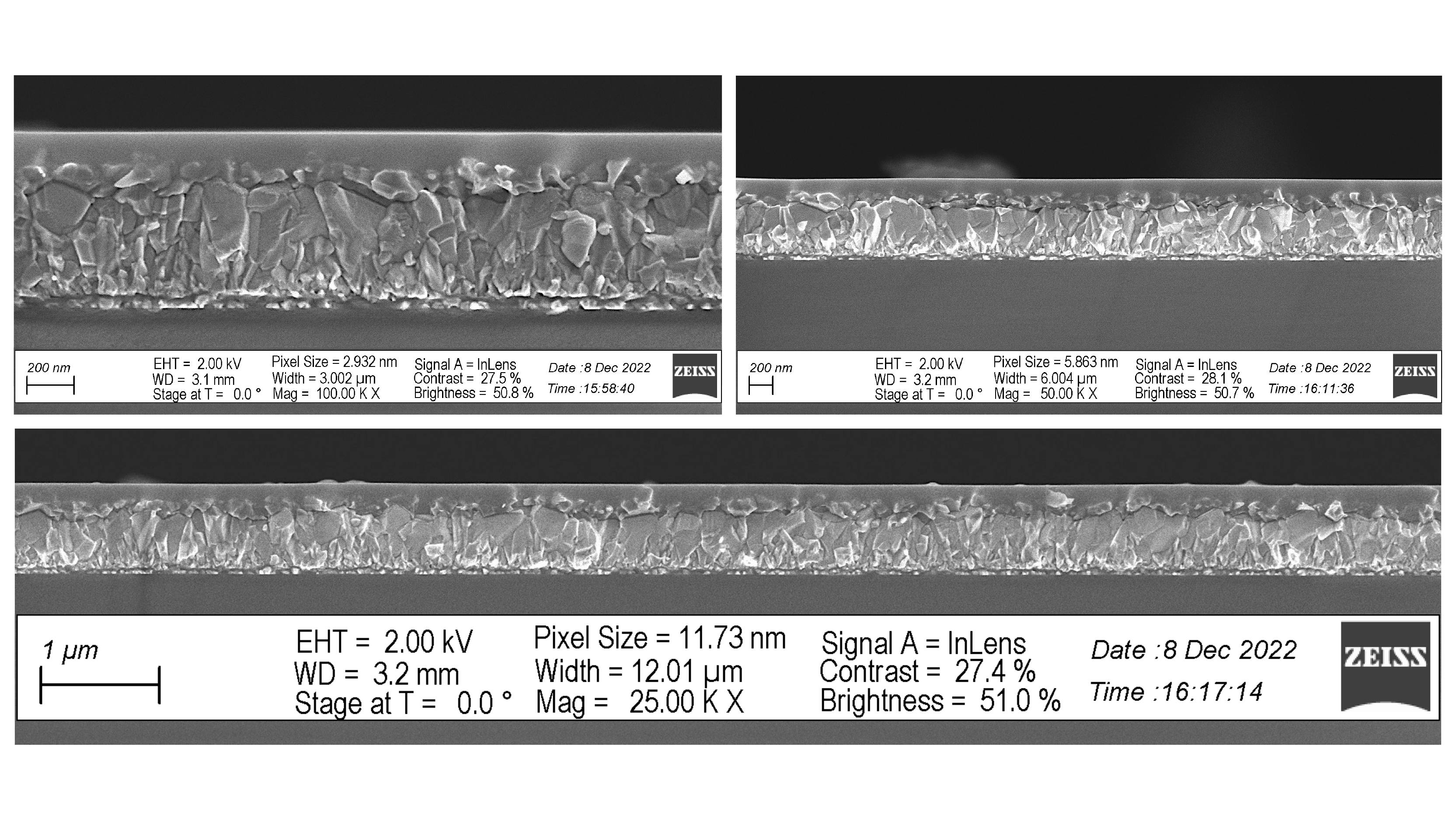}
    \caption{Additional cross-sectional SEM images of devices immediately after superstrate laser-annealing at different magnifications. The images show the formation of a buried layer of selenium crystallites.}
    \label{fig:ESI11}
\end{figure*}

\begin{figure*}[ht!]
    \centering
    \includegraphics[width=\textwidth,trim={0 250 0 0},clip]{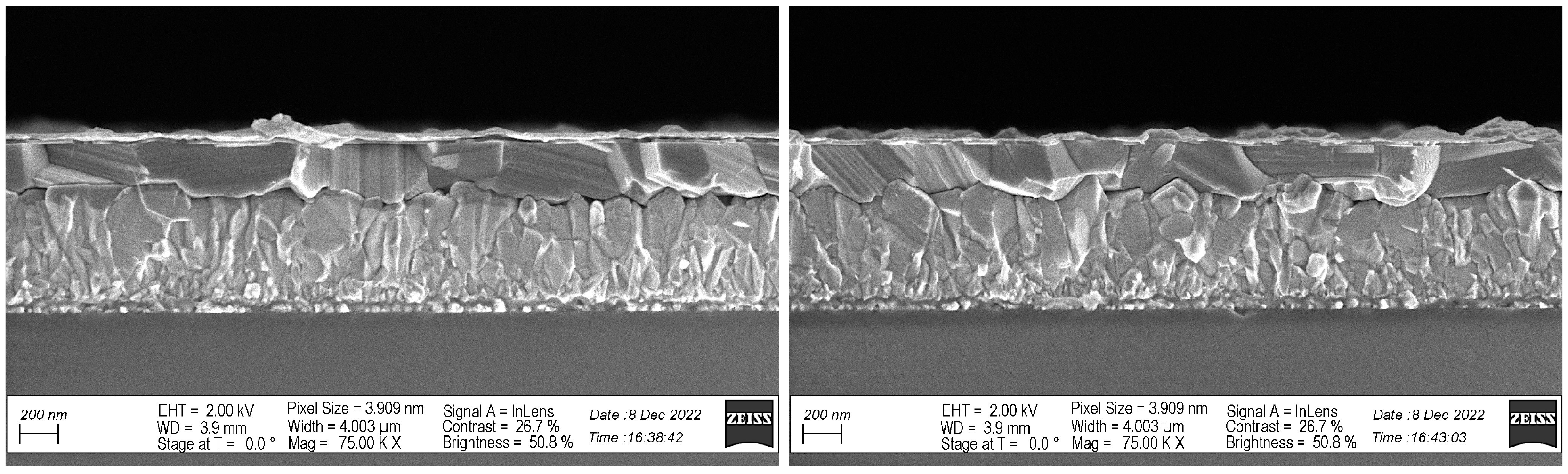}
    \caption{Additional cross-sectional SEM images of devices where the photoabsorber is solid-phase epitaxially grown from a buried seed layer formed using superstrate laser-annealing.}
    \label{fig:ESI4}
\end{figure*}

\begin{figure*}[ht!]
    \centering
    \includegraphics[width=\textwidth,trim={0 0 0 0},clip]{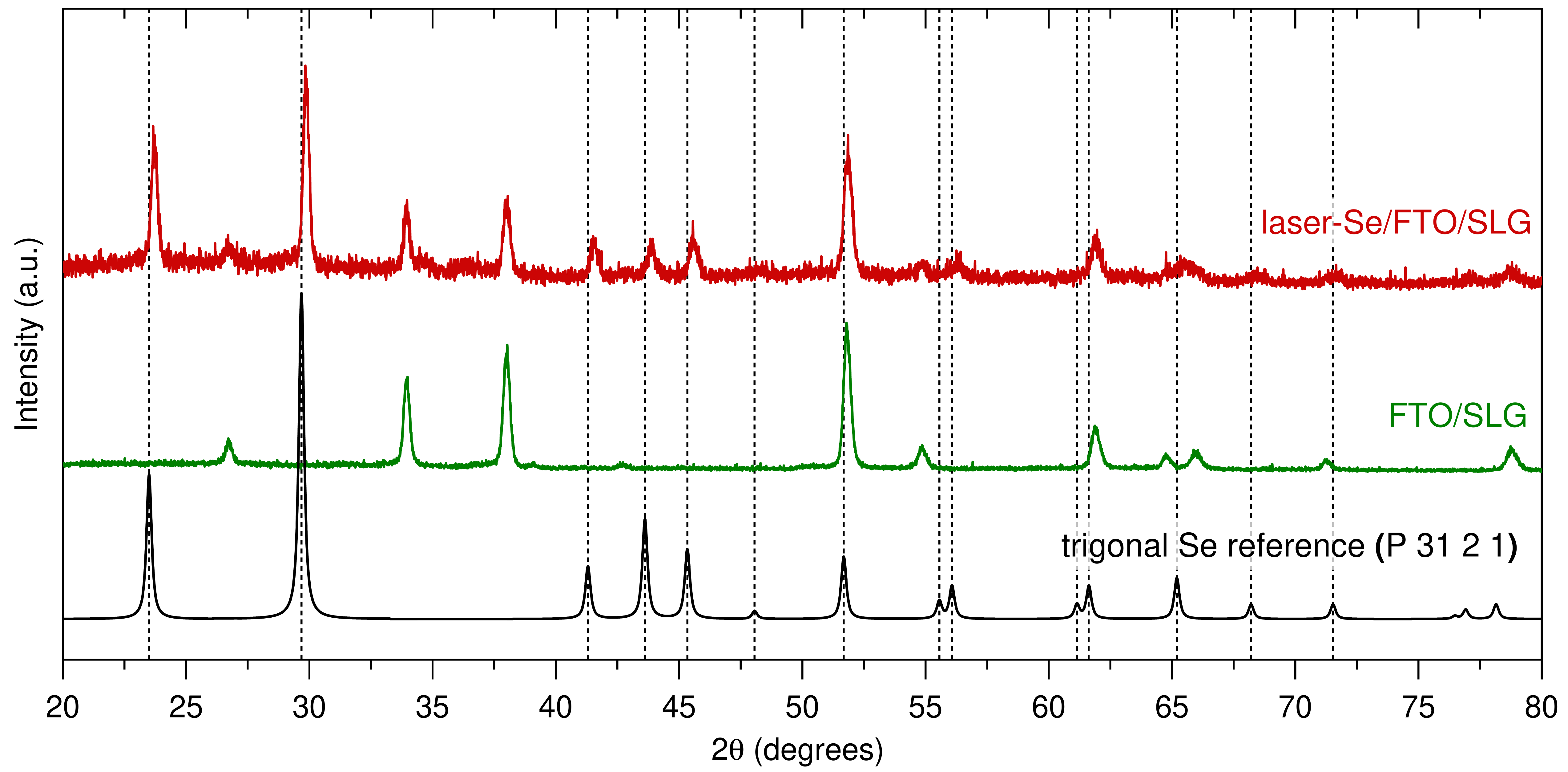}
    \caption{Grazing-incidence X-ray diffraction (GIXRD) pattern of the substrate laser-annealed sample from Figure 1 in the main text. Even though the integrity of the surface is compromised, the presence of trigonal selenium is observed.}
    \label{fig:ESI5}
\end{figure*}

\begin{figure*}[ht!]
    \centering
    \includegraphics[width=\textwidth,trim={0 0 0 0},clip]{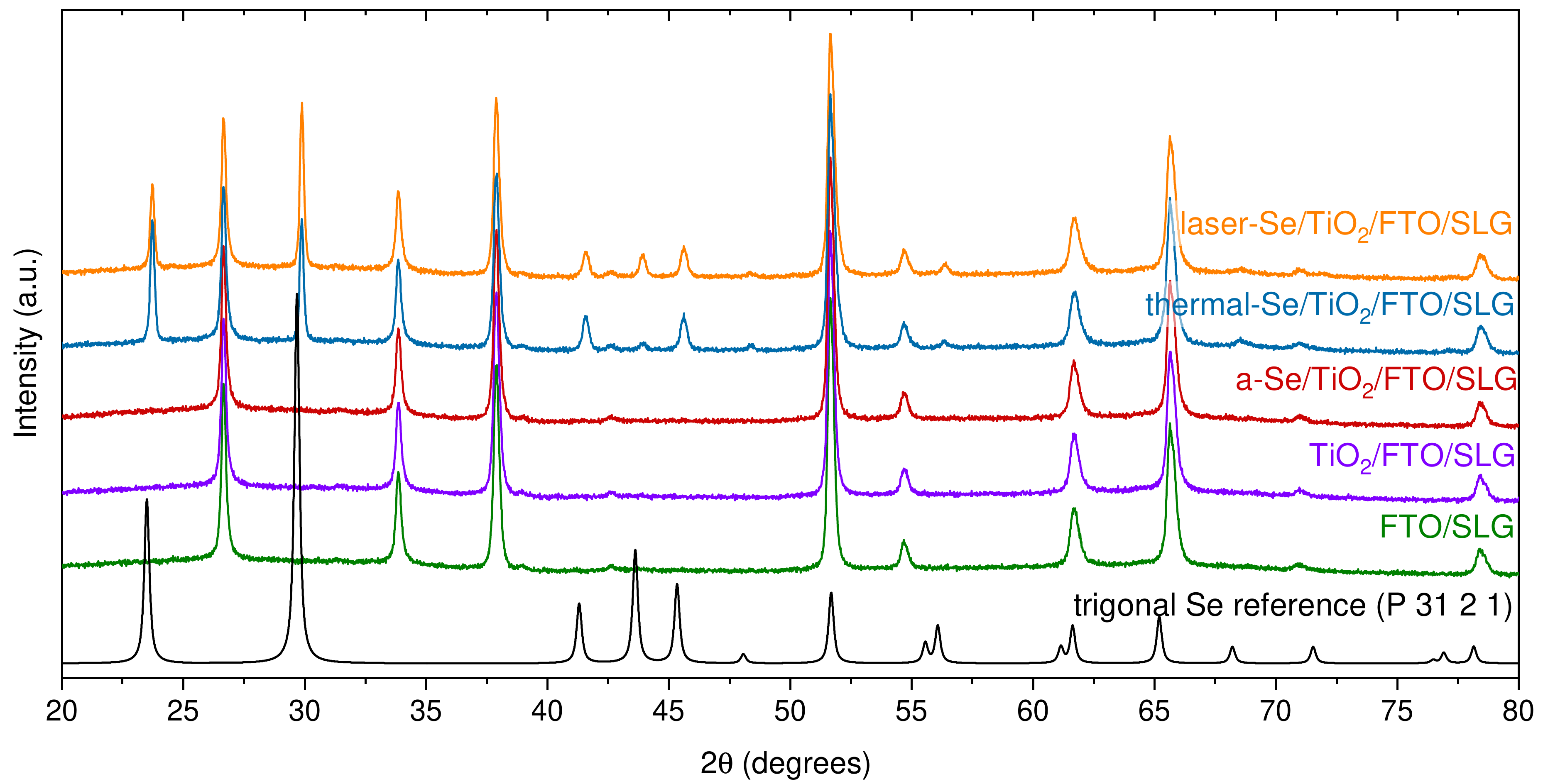}
    \caption{Grazing-incidence X-ray diffraction (GIXRD) patterns of all layers in the device stack until the absorber has been crystallized using either superstrate laser-annealing or thermal annealing.}
    \label{fig:ESI6}
\end{figure*}

\begin{figure*}[ht!]
    \centering
    \includegraphics[width=0.86\textwidth,trim={0 0 0 0},clip]{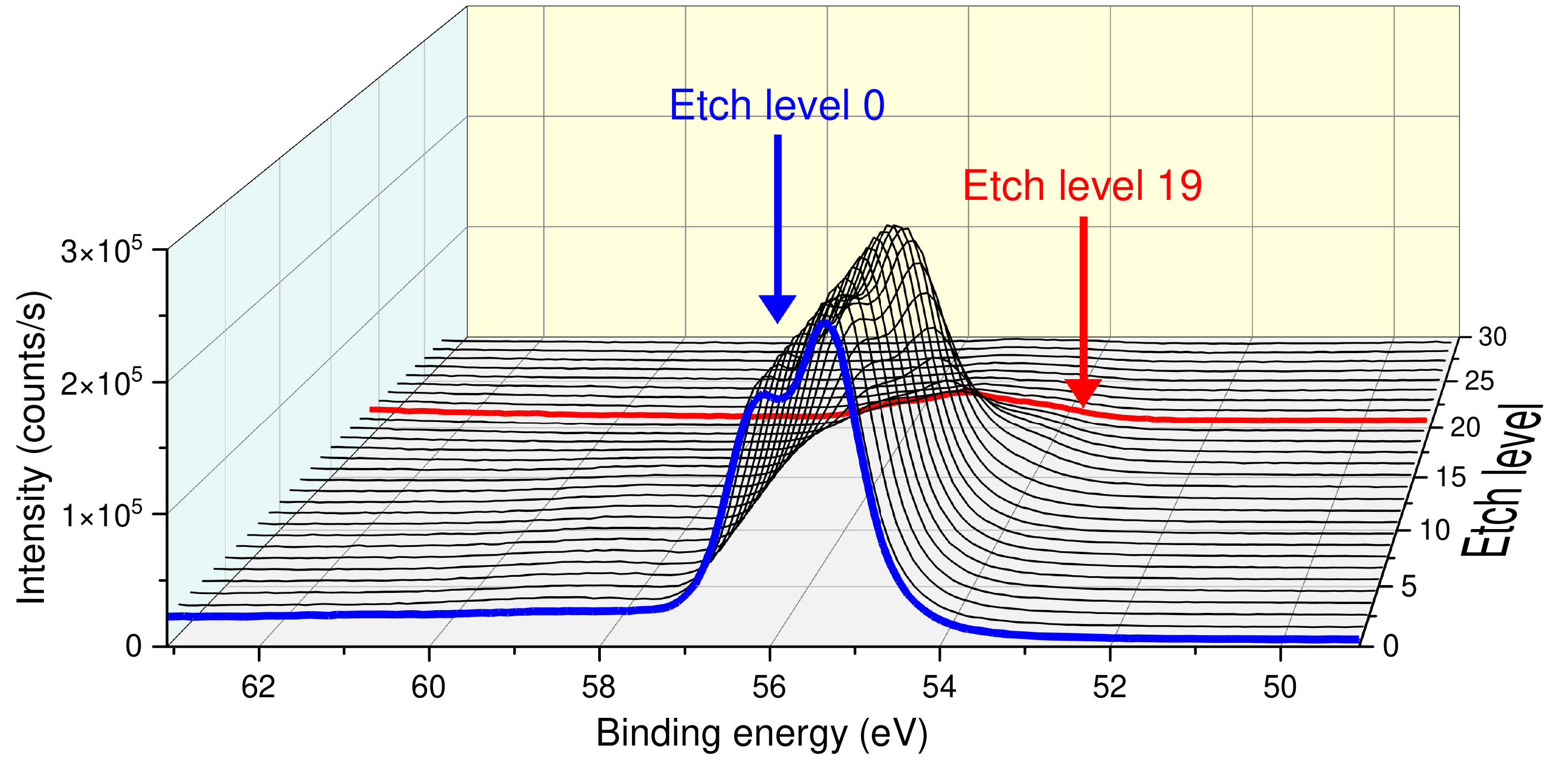}
    \includegraphics[width=\textwidth,trim={0 0 0 0},clip]{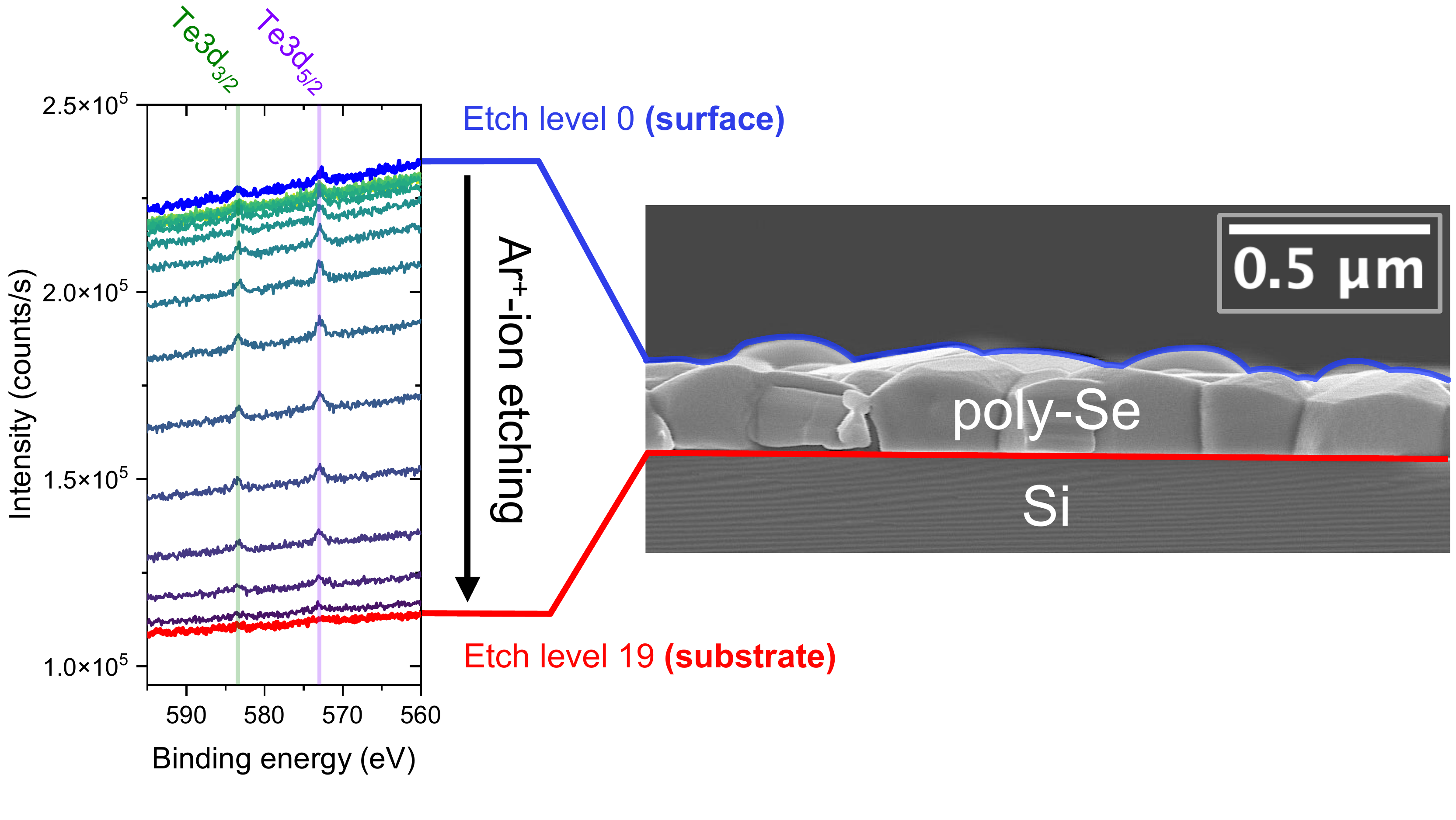}
    \caption{XPS depth profile and SEM-cross-section of thermally crystallized poly-Se on silicon. The etch level at which we have sputtered through the 300 nm photoabsorber is determined from core level scans of selenium. As the 3d-orbital doublet peaks of tellurium are detected at all etch levels from the surface to the silicon substrate, the tellurium adhesion layer must have interdiffused into the bulk of the selenium thin-film.}
    \label{fig:ESI3}
\end{figure*}
\newpage

\begin{figure*}[ht!]
    \centering
    \includegraphics[width=\textwidth,trim={0 0 0 0},clip]{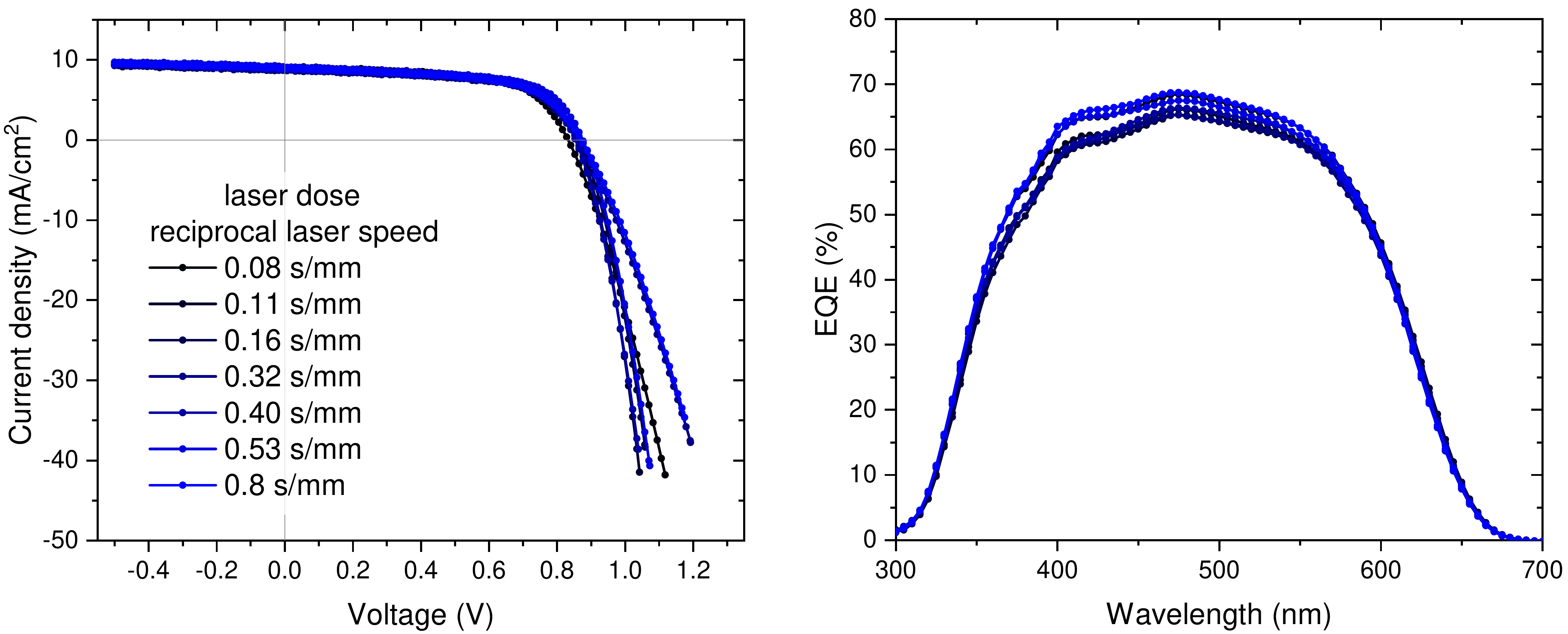}
    \caption{Current-voltage (\textit{J-V}) curves and external quantum efficiency (EQE) spectra of final devices after the post-annealing processing step. Regardless of laser dose, all devices perform quite similarly, and no observable trends as a function of increasing dose have been observed.}
    \label{fig:ESI12}
\end{figure*}

\begin{figure*}[ht!]
    \centering
    \includegraphics[width=\textwidth,trim={0 0 0 0},clip]{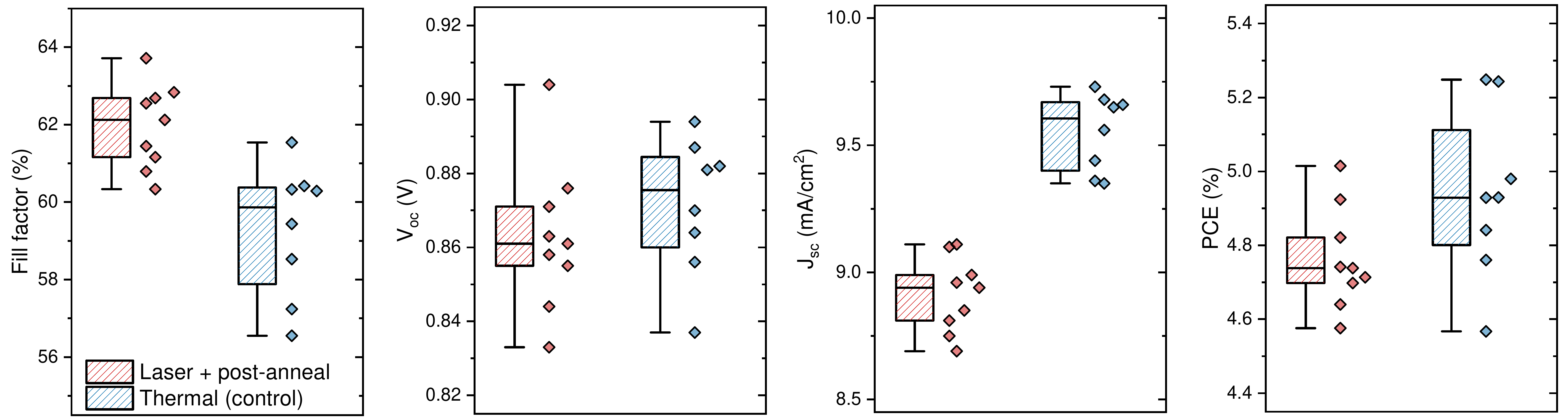}
    \caption{Photovoltaic device performance metrics of the batch of laser-annealed samples after the post-annealing step compared to the batch of thermally annealed control devices.}
    \label{fig:ESI2}
\end{figure*}

\begin{figure*}[ht!]
    \centering
    \includegraphics[width=0.45\textwidth,trim={0 0 0 0},clip]{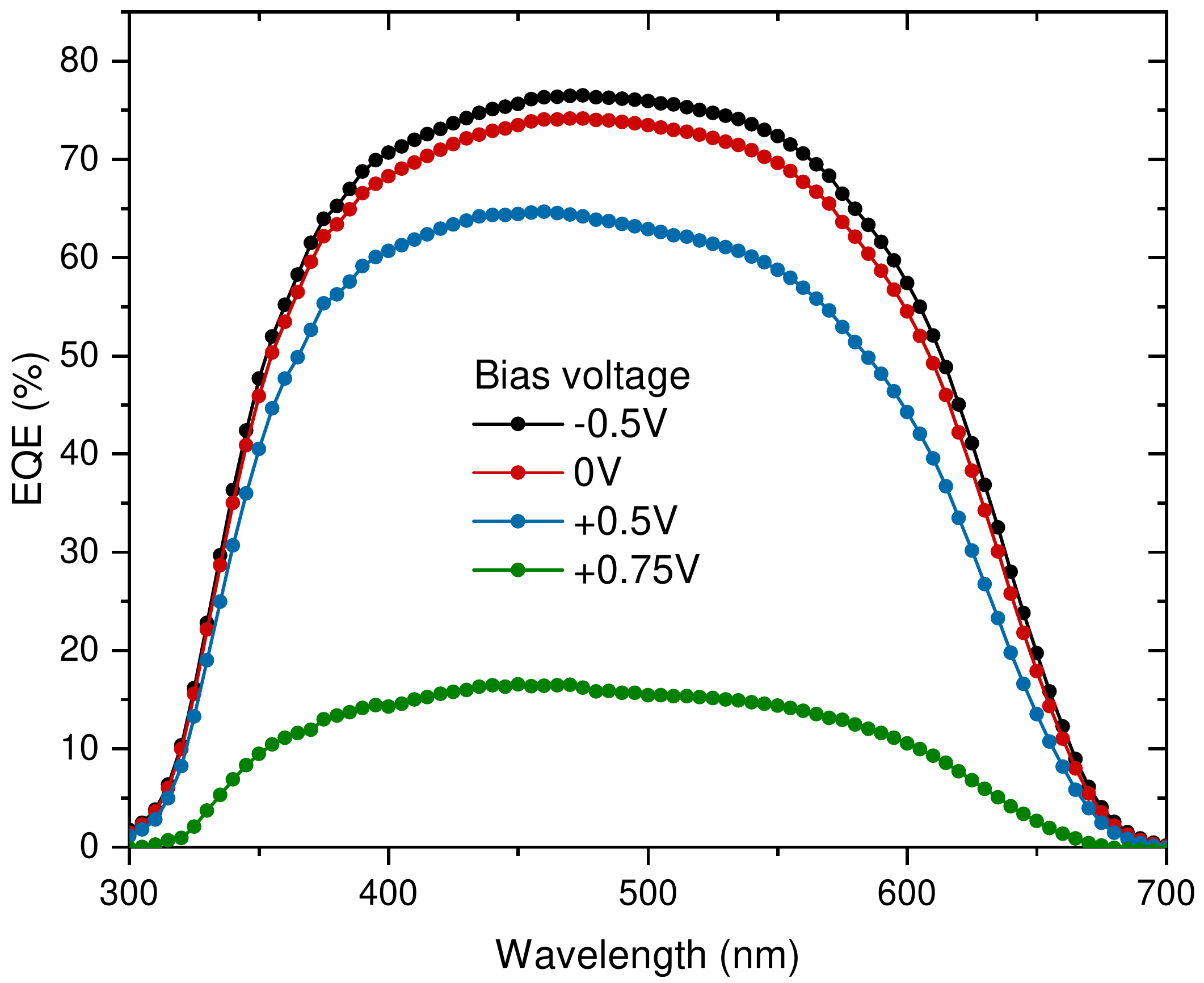}
    \caption{External quantum efficiency (EQE) spectra of a superstrate laser-annealed sample at various externally applied voltages.}
    \label{fig:ESI8}
\end{figure*}\newpage

\begin{figure*}[t!]
    \centering
    \includegraphics[width=\textwidth,trim={0 0 0 0},clip]{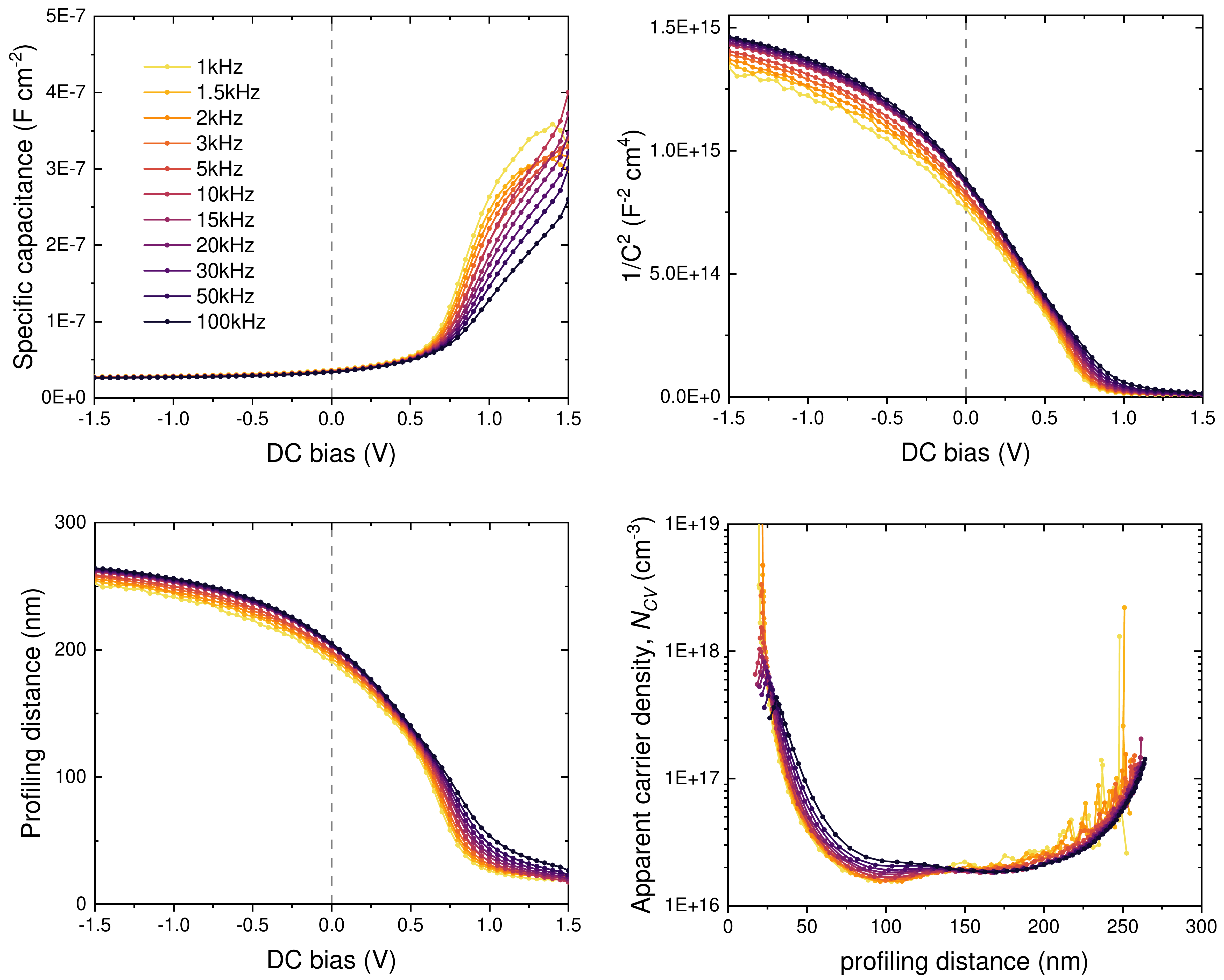}
    \caption{C-V measurements of a laser-annealed solar cell, where the diode capacitance is analyzed using the diode depletion approximation at various frequencies.}
    \label{fig:ESI13}
\end{figure*}